\documentclass[11pt]{article}

\usepackage{amsmath, amssymb, amsthm, amsbsy, ifthen, graphicx}

\usepackage{authblk}

\usepackage{floatrow}
\usepackage[label font={bf,Large},labelformat=simple]{subfig}
\usepackage{xr}

\usepackage[utf8]{inputenc}
\usepackage{fullpage}
\usepackage{amsmath}
\usepackage{color}
\usepackage{graphicx}
\usepackage{breqn}
\usepackage{algorithm}
\usepackage{algpseudocode}
\usepackage[hypertexnames=false]{hyperref}
\usepackage{natbib}

\renewcommand{\d}{\mathbf{d}}
\newcommand{\0}{\mathbf{0}}
\newcommand{\1}{\mathbf{1}}
\allowdisplaybreaks
\newcommand{\beginsupplement}{%
        \setcounter{table}{0}
        \renewcommand{\thetable}{S\arabic{table}}%
        \setcounter{figure}{0}
        \renewcommand{\thefigure}{S\arabic{figure}}%
        \setcounter{section}{0}
        \renewcommand{\thesection}{S\arabic{section}}%
     }
\usepackage[utf8]{inputenc}
\usepackage{fullpage}
\usepackage{amsmath}
\usepackage{color}
\usepackage{graphicx}
\usepackage{amssymb}
\usepackage{matlab-prettifier}
\usepackage{bm}
\usepackage{multirow}
\usepackage{caption}
\usepackage{natbib}

\usepackage{caption}

\providecommand{\keywords}[1]
{
  \small	
  \textbf{\textit{Keywords---}} #1
}

\usepackage{color}   
\usepackage{hyperref}
\hypersetup{
    colorlinks=true, 
    linktoc=all,     
    linkcolor=blue,  
}



\begin{document}
\date{}
\title{\vspace{-1cm}Approximate Bayesian computation for Markovian binary trees in phylogenetics}

\author[1,2]{\small Mingqi He}
\author[1]{\small Sophie Hautphenne}
\author[1,2]{\small Yao-ban Chan}

\affil[1]{\footnotesize School of Mathematics and Statistics, The University of Melbourne}
\affil[2]{\footnotesize Melbourne Integrative Genomics, The University of Melbourne}

\maketitle

\begin{abstract}
    Phylogenetic trees describe the relationships between species in the evolutionary process, and provide information about the rates of diversification. To understand the mechanisms behind macroevolution, we consider a class of multitype branching processes called Markovian binary trees (MBTs). MBTs allow for trait-based variation in diversification rates, and provide a flexible and realistic probabilistic model for phylogenetic trees. 

    We develop an approximate Bayesian computation (ABC) scheme to infer the rates of MBT parameters by exploiting the information in the shapes of phylogenetic trees.
    We evaluate the accuracy of this inference method using simulation studies, and find that our method is able to detect variation in the diversification rates, with accuracy comparable to, and generally better than, likelihood-based methods. In an application to a real-life phylogeny of squamata, we reinforce conclusions drawn from earlier studies, in particular supporting the existence of ovi-/viviparity transitions in both directions. 

    Our method demonstrates the potential for more complex models of evolution to be employed in phylogenetic inference, in conjunction with likelihood-free schemes.
\end{abstract}


\section{Introduction}

The rates of species generation and extinction, called diversification rates, are reflected in the evolutionary history of the species family, which can be represented in the form of a phylogenetic tree. A large number of statistical models have been developed to model such trees \citep{morlon2014phylogenetic}. These typically take the form of birth-death models, which are fitted to trees reconstructed from DNA or protein sequence data. By then inferring the parameters of the model, we can gain information about the diversification rates of the species.

The traditional model used in phylogenetics is the linear (or constant-rate) birth-death process, where all species are considered identical and independent, and speciate and go extinct at rates that are constant through time. Although computationally tractable, this is an oversimplification of reality: both speciation and extinction rates can be influenced by many different factors, such as geographical location \citep{hillebrand2004generality}, body size \citep{gittleman1998body}, mating systems \citep{barraclough1998revealing}, and dietary requirements \citep{vellend2011measuring}. Identifying and quantifying the variation in these rates between species can provide crucial insights into macroevolutionary dynamics \citep{stadler2013recovering}.

To model variation in diversification rates more realistically, it is necessary to replace the linear birth-death process by a more complex model that can account for these variations. One way to do this is to consider discrete traits of the species, known as \emph{states}, which may influence the diversification rates. States may represent a number of different phenotypes, as discussed above, or may simply provide more flexibility in modelling without necessarily corresponding to particular physical traits.

The \textit{binary state speciation and extinction} (BiSSE) model \citep{maddison2007estimating} allows each species to have two such states, which we assume to be observed in extant species. For each state, the rates of speciation and extinction are constant, as are rates of transition from one state to the other. In general, the children of a speciation are assumed to be in the same state as the parent. \citet{maddison2007estimating} used systems of differential equations to calculate the likelihood of a given phylogenetic tree under this model; this allows an estimation of diversification rates using maximum likelihood methods. This approach has also been applied to a variety of related models, including models that allow for multiple states (MuSSE, \citep{fitzjohn2009estimating}), two hidden states (HiSSE, \citep{beaulieu2016detecting}), and so-called ``several examined and concealed" states (SecSSE, \citep{herrera2019detecting}). More recently, \citet{louca2020general} introduced a fast method of calculating likelihoods under the BiSSE model, allowing much larger trees to be analysed.

The BiSSE model is a special case of a tractable class of Markovian continuous-time branching processes called the \emph{Markovian binary tree} (MBT, \citep{kontoleon2006markovian,hautphenne2012markovian,hautphenne2014algorithm}). In this model, species can be in one of any number of \emph{phases} (equivalent to states), and each phase has specific (but constant throughout time) rates of birth (into children of given phases, not necessarily identical to the parent or to each other), transition (into a different phase), and death. This model provides the necessary flexibility to account for variation in diversification rates. However, as the number of phases increases, the model becomes more complex, and computing the likelihood of phylogenies using a system of differential equations becomes increasingly computationally expensive. Another drawback of likelihood methods is their strong dependence on the inferred phylogeny, including branch lengths, which cannot always be reliably estimated from sequence data.

We approach the problem of estimating diversification rates under the MBT model by using the fact that these models are easy to simulate. This enables us to use \textit{approximate Bayesian computation} (ABC, \citep{beaumont2002approximate}), a likelihood-free method. In this method, we simulate a number of observations from a generative prior model (in this case the MBT with parameters drawn from a prior distribution), and compare these observations to the real data by means of various summary statistics. The simulated observations that are `close enough' to the real data (based on a tolerance level) form empirical distributions for the parameters which are approximations to their posteriors. Because ABC relies only on the summary statistics calculated from the observed trees, it is less reliant on local features (such as individual branch lengths) than a full likelihood method.

ABC is a well-established statistical method, and numerous methodological developments have been made in recent years. One such advancement is \textit{ABC Population Monte Carlo} (ABC-PMC, \citep{beaumont2009adaptive}), which uses an iterated approach where the approximate posterior distributions form the basis of the prior distributions for the next iteration. When combined with decreasing tolerance levels, this allows for gradual convergence to more accurate posteriors.

ABC methods have been used previously to infer diversification rates for linear birth-death models \citep{bokma2010time,janzen2015approximate,rabosky2008explosive}. In particular, \citet{janzen2015approximate} examined the performance of summary statistics for use in ABC methods, finding that traditional statistics such as phylogenetic diversity or tree size often perform poorly. They introduced the normalised lineage-through-time (nLTT) curve as a statistic to improve the performance of ABC methods. However, as far as we know, no one has yet attempted to perform ABC with a more general and realistic model.

In this paper, we apply ABC-PMC to infer diversification rates from phylogenetic trees under the MBT model
. We focus on cases where the MBT has two phases, and where it is either possible to transition only from one phase to the other (reducible case), or in either direction (irreducible case). We develop and test various summary statistics tailored to these scenarios, enabling us to accurately infer the diversification rates, particularly in the reducible case. In simulations, our method shows higher accuracy compared to the maximum-likelihood method of \citep{louca2020general}. We further demonstrate the applicability of our method by applying it to a real dataset of squamata (reptiles), where the phases represent the method of bearing offspring (egg-laying or live-bearing); our analysis yields inferred parameters similar to earlier studies based on the BiSSE model.
This highlights the potential of ABC and MBTs as a versatile approach for inferring diversification rates in phylogenetics.

\section{Methods}

\subsection{Inference of Markovian binary tree parameters using approximate Bayesian computation}

Markovian binary trees (MBTs) are a flexible class of continuous-time branching processes, where each individual in a population (here, each species in a family) exists in one of $n$ phases. In this paper, we consider only the simplest case $n = 2$. An individual in phase $i$ can:
\begin{itemize}
	\item transition to phase $j\neq i$ at constant rate $q_{ij}$;
	\item give birth to a child in the same phase $i$ at constant rate $\lambda_i$ (remaining in the same phase); and
	\item die at constant rate $\mu_i$.
\end{itemize}
The asymptotic behaviour of the population size is exponentially controlled by the \emph{growth rate} $\omega$, which can be calculated from these parameters; see Supplementary Section~\ref{sec:mbt} for more details, and for a more general formulation of the MBT process.

To infer the diversification rates of an MBT model from a set of $m$ observed phylogenetic trees, we apply the ABC-PMC method \citep{beaumont2009adaptive}. This method proceeds over a series of iterations; at each iteration a standard ABC method is applied where parameters are proposed from a prior distribution, a dataset is simulated using the proposed parameters, and the parameters are accepted if the simulated dataset is `similar' to the observed dataset when comparing various summary statistics. 
In the ABC-PMC method, the accepted parameters in one iteration form the basis for the prior in the next iteration, allowing the posteriors to converge towards the true values. We refer to Supplementary Section \ref{sec:abc-pmc} for more details.

\subsection{Summary statistics}\label{sec:summary}

The effectiveness of the ABC-PMC method crucially relies on the careful selection of an appropriate suite of summary statistics to compare the observed and simulated trees.
These statistics must collectively be able to capture variation in each of the parameters of the model. The statistics we use are as follows.
\begin{enumerate}
	\item \textbf{Average branch length}.
	\item \textbf{Tree height} (time to most recent common ancestor). Note that our simulated trees have a fixed number of extant species, but variable height. In situations where the tree height is fixed, the number of extant species may be used instead.
	\item \textbf{Normalised lineage-through-time (nLTT) curve} \citep{janzen2015approximate}. This curve shows the growth of the number of species against time, normalised by both the total time and the total number of extant species. To compute the distance between two nLTT curves, we calculate the absolute area between them:
		\[\Delta\text{nLTT} = \int_0^1 | \text{nLTT}_1(t) - \text{nLTT}_2(t) |\,dt.\]

		We set the initial tolerance value for this  statistic to be $0.05$. Due to the computational time required to generate a sample that can provide a close enough nLTT curve, we only use this statistic after the 20th iteration.
	\item \textbf{Colless balance index} \citep{colless1982phylogenetics}. For a tree $T$, this is defined as the sum over all internal nodes $v \in V_{\text{int}}(T)$ of the absolute difference between the number of left and right descendant leaves. In other words, if $v$ has children $\{v_l,v_r\}$, and $\kappa_u$ is the number of descendant leaves from node $u$, then the Colless index of $T$ is
		\[C(T) = \sum_{v \in V_{\text{int}}(T)} |\kappa_{v_l} - \kappa_{v_r}|.\]
	\item \textbf{Balance index for each phase}. To capture imbalance in the shape of the tree in each phase, we propose a novel extension of the Colless balance index to each phase; if $\kappa^{(i)}_u$ is the number of descendant leaves in phase $i$ from node $u$, we define the balance index for phase $i$ as
		\[C_i(T) = \sum_{v \in V_{\text{int}}(T)} |\kappa_{v_l}^{(i)} - \kappa_{v_r}^{(i)}|.\]
	\item[R1.] \textbf{Tree height for trees that contain phase-1 leaves} (\textit{reducible case only}). In the reducible case, species in phase 2 cannot transition to phase 1. Thus a tree with some leaves in phase 1 indicates either a large birth rate for phase 1, or a small transition rate from phase 1 to phase 2. These two situations are distinguished by the overall tree height; thus, we use the tree height for the subset of trees that contain at least one leaf in phase 1.
	\item[R2.] \textbf{Average branch length above
 phase-1 leaves} (\textit{reducible case only}). In the reducible case, all branches above leaves in phase 1 must themselves be in phase 1. Thus we use the average length of these branches as a summary statistic, which captures the birth and death rates for phase 1.
	\item[I1.]  \textbf{Proportion of phase-1 leaves} (\textit{irreducible case only}). The proportion of leaves in phase 1 captures the variation between the growth rates of phase-1 lineages versus phase-2 lineages.
	\item[I2.]  \textbf{Transition statistic for each phase} (\textit{irreducible case only}). We propose the \emph{transition statistic} of a tree $T$ for phase $i$ as
		\[S_i(T) = \sum_{v \in V'_{\text{int}}(T)} - \frac{\ln P_v^i}{d_v + \gamma}, \qquad i = 1,2,\]
		where $V'_{\text{int}}(T)$ is the set of internal nodes with descendant leaves in both phases, $P_v^i$ is the proportion of phase-$i$ leaves in the descendant leaves of $v$, $d_v$ is the time of node $v$, and $\gamma$ is a small constant (here set to $0.02$). 

        This statistic can be intuitively justified by considering that its dominant terms correspond to subtrees with small height (small $d_v$). We can approximate such a subtree by considering the paths from the root to each leaf as independent; if we assume that the root of the subtree is not in phase $i$, then the proportion of leaves in phase $i$ is approximately the probability of a transition to phase $i$ on a branch of length $d_v$, i.e., $e^{-q_{.i} d_v}$. Inverting this formula for the transition rate to phase $i$ gives the transition statistic for phase $i$. (We add a small constant $\gamma$ to the denominator to avoid extremely small $d_v$ values from dominating the statistic.)

\end{enumerate}

\section{Results}

We first study the accuracy of our method through simulations.

\subsection{Reducible case}\label{sec:eq_dr}

We first consider the case where the MBT is reducible and the death rates are equal, i.e., $q_{21} = 0, \mu := \mu_1 = \mu_2$. This results in four free parameters for the MBT model. We only consider supercritical processes (i.e., with growth rate $\omega>0$), so that the expected number of lineages grows without bound.
Our observed dataset $Y_{\text{obs}}$ consists of 100 trees, each with 50 leaves. These trees are generated from the MBT model with the extinct subtrees removed.

In the ABC-PMC algorithm, we set the number of simulated trees for iteration $t$ as follows:
\[n_t = \left\{ \begin{array}{rl} 1, & \qquad t = 1, \\ 10, &  \qquad t = 2,\dots,10, \\ 100, & \qquad t = 10,\dots,30. \end{array} \right.\]
For earlier iterations, this enables us to quickly explore the parameter space and accept the required number of samples, which we set to $N = 200$. For later iterations, we simulate (as required) datasets that are the same size as the observed dataset.

We first consider a `default' set of parameters $(\lambda_1, \lambda_2, \mu, q_{12}) = (1.5, 0.51, 0.15, 0.5)$, which corresponds to a growth rate of $\omega = 0.85$. These values are chosen to ensure that there is a noticeable difference between the birth rates of the two phases; we expect some parts of the trees will be in the fast-growing phase 1, while others will be in the slower-growing phase 2, creating a detectable imbalance.

In Supplementary Figure~\hyperlink{text:4ptrace}{S11a}, we show the posterior means at each iteration for one run of the algorithm. We see that the posterior means converge towards the true values as the number of iterations increases.
The final posterior distributions are shown in Supplementary Figure~\hyperlink{text:4ppostmodall1}{S12}; here we see that the distributions are closely concentrated around the true values. For all four parameters, the final posterior means are all within 4\% of the true values. This suggests that the algorithm has achieved good convergence and can accurately recover the true parameter values.

Of particular significance is the capability to infer the true growth rate $\omega$ of the MBT process. 
In Supplementary Figure~\hyperlink{text:4ptrace}{S11b}, we illustrate (for a single run) the convergence of our estimates towards the true value as the number of iterations increases. Although the estimate is quite accurate, a minor positive bias is noticeable, consistently observed across different runs of the algorithm. This might be attributed to the fact that observed trees are conditioned on having 50 leaves (and hence are not extinct), resulting in a slightly higher effective growth rate.

\begin{figure}[htp]\centering
	\includegraphics[width=0.8\textwidth]{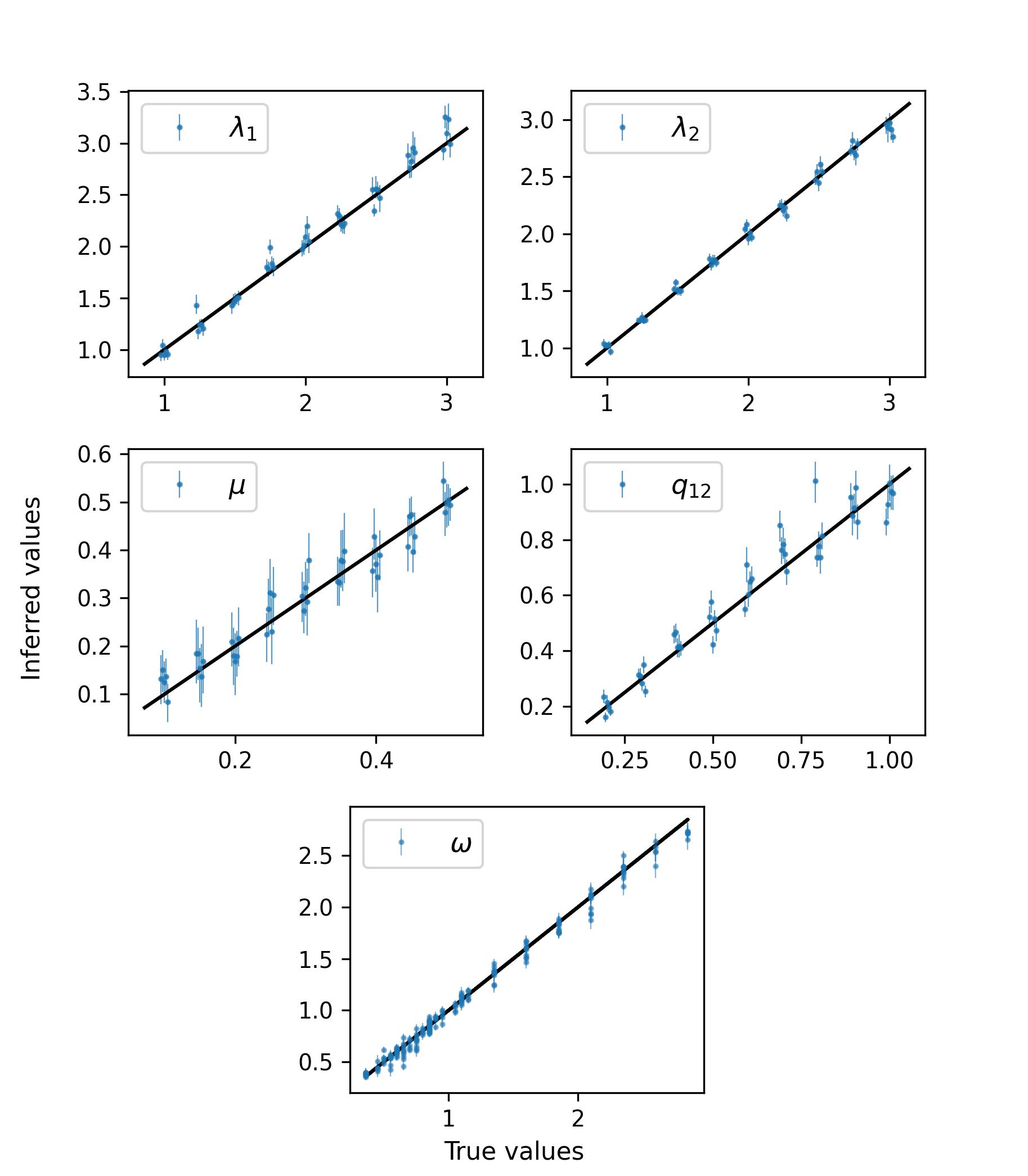}
	\caption{Inferred vs true parameter values for the reducible case with equal death rates. Each parameter value is varied while keeping the other parameters fixed at their default values. Error bars represent the 50\% credible intervals for the final posteriors.}
	\label{fig:4pars}
\end{figure}

We tested the accuracy of the method by varying the true parameters one at a time over a range of values, and repeated the inference (inferring all parameters) for 5 runs at each value. The results in Figure~\ref{fig:4pars} show accurate inference of the true parameter values in all cases.


For most parameters, the accuracy of inference decreases as the parameter value increases (e.g., the 50\% credible intervals become larger). The exception is the death rate $\mu$, where the accuracy appears relatively unaffected by the parameter value.

Next, we consider the general reducible case, where the death rates may differ between the two phases, resulting in five free parameters. The extra degree of freedom makes inference slightly more difficult. We consider the default parameters $(\lambda_1, \lambda_2, \mu_1, \mu_2, q_{12}) = (1.5, 0.51, 0.7, 0.15, 0.5)$, where the parts of the tree in phase 1 both speciate and go extinct relatively quickly, while parts of the tree in phase 2 speciate and go extinct relatively slowly, and have an overall lower growth rate. 

In Supplementary Figures~\hyperlink{text:5ptrace}{S13} and \hyperlink{text:posterior5p}{S14}, we show the trace plots and final posterior distributions for a single run. Similar to the equal death rate case, the posterior distributions show good convergence by the 30th iteration. 
Our inference is highly accurate for most parameters, with slight reductions in accuracy observed for $\mu_1$ and $q_{12}$. This decrease in accuracy could be attributed to the reducible nature of the process; as most phase-1 lineages eventually transition to phase 2, there is less information available in the data regarding the phase-1 death and transition rates.
When assuming equal death rates, it is possible to infer the phase-1 death rate from phase-2 data,
but this is no longer possible here. Nonetheless, our inference of the overall growth rate ($\omega=0.36$) remains accurate (Supplementary Figure \hyperlink{text:5ptrace}{S13d}).

\begin{figure}[htp]\centering
	\includegraphics[width=0.8\textwidth]{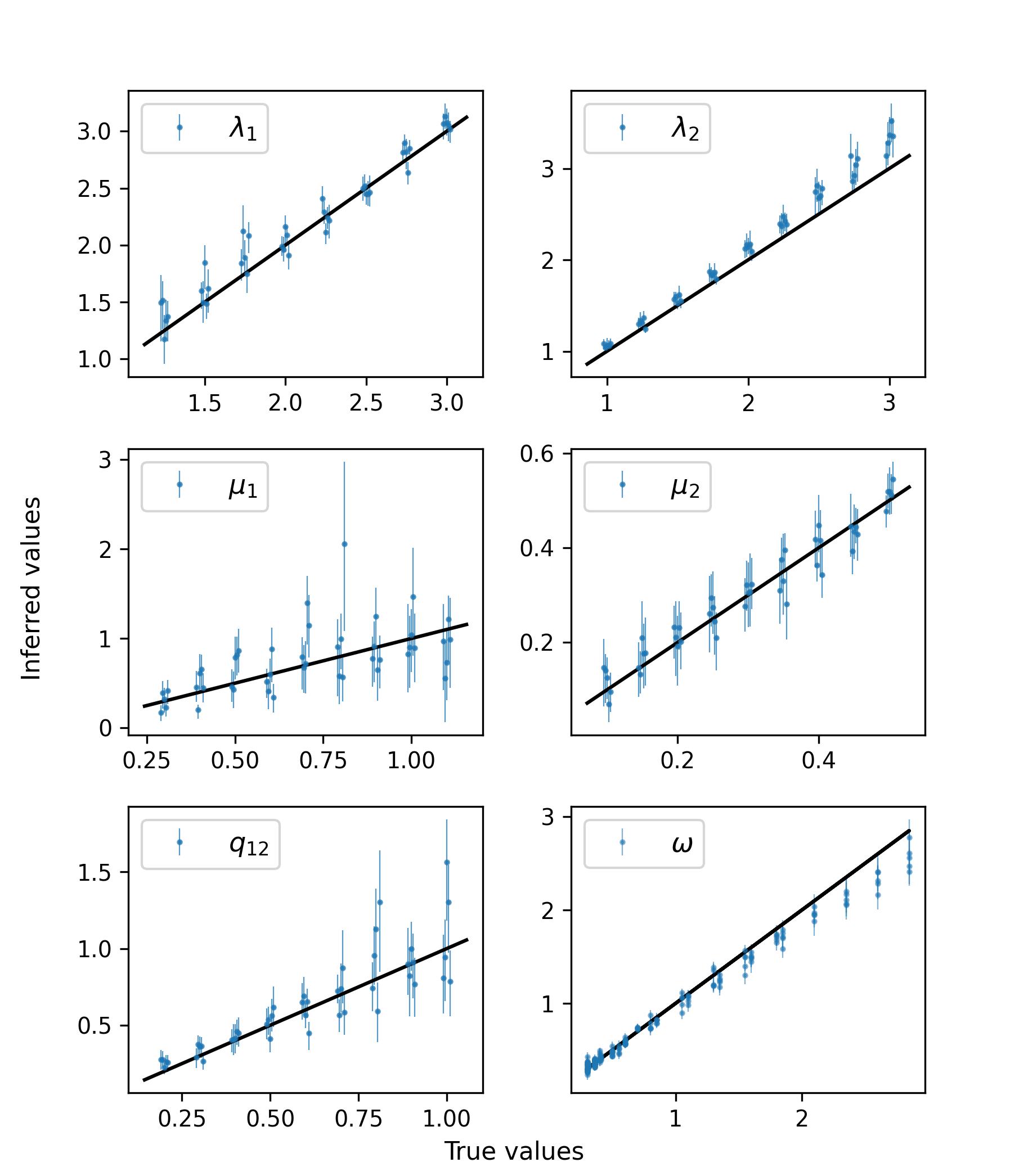}
	\caption{Inferred vs true parameter values for the reducible case with arbitrary death rates. One parameter value is changed at a time, while the other parameter values stay fixed at their default values. Error bars represent the 50\% credible intervals for the final posteriors.}
	\label{fig:5pars}
\end{figure}

As above, 
we varied the true parameters one at a time over a range of values and repeated the inference 5 times for each parameter setting. 
The results, shown in Figure~\ref{fig:5pars}, indicate accurate inference, although there are some instances with noticeable errors. 

As with the equal death rate case, the estimation of the phase-2 birth rate $\lambda_2$ and transition rate $q_{12}$ become less accurate as the parameter value increases, while the accuracy of the phase-2 death rate $\mu_2$ is largely unaffected by the parameter value. However, here there is a small positive bias for the phase-2 birth rate which increases with the parameter. In contrast, the estimation of the phase-1 birth rate $\lambda_1$ becomes more accurate as the parameter value increases, while the phase-1 death rate $\mu_1$ becomes less accurate. This can again be explained by the reducible nature of the process; the larger the ratio of phase-1 birth rate to death rate, the longer the tree spends in phase 1 before most lineages transition eventually to phase 2. This gives more information for phase 1, leading to better estimation of those rates.

\subsection{Irreducible case}\label{sec:irred}

We now consider the irreducible case, where transitions in both directions are allowed. We also allow death rates to differ between phases, resulting in six free parameters. For this case, we use the I1 and I2 summary statistics in place of the R1 and R2 statistics from Section \ref{sec:summary}. The remainder of the method is unchanged.

In order to verify that the summary statistics used are sensitive to the parameters of the MBT process, we visualise (in Supplementary Section \ref{sec:sensitivity}) the variation of the summary statistics as the MBT parameters are varied. The statistics appear to have the necessary sensitivity for inference in the irreducible case.

In Supplementary Figure \hyperlink{text:6ptrace}{S15}, we show the trace plots for a single run for the default parameters $(\lambda_1,\lambda_2,\mu_1,\mu_2,q_{12},q_{21}) = (3,2,1,0.5,0.5,0.25)$. Again, it is clear that the process has converged by the end of the 30th iteration. In Supplementary Figure~\hyperlink{text:posterior6p}{S16}, we show the final posterior distribution. As in the general reducible case, the resulting posterior distributions are more dispersed for the death rates  than the birth rates. Overall, our inference has high accuracy for all parameters, indicating the summary statistics used can detect the influence of different phases. 

\begin{figure}[htp]\centering
    \includegraphics[width=0.8\textwidth]{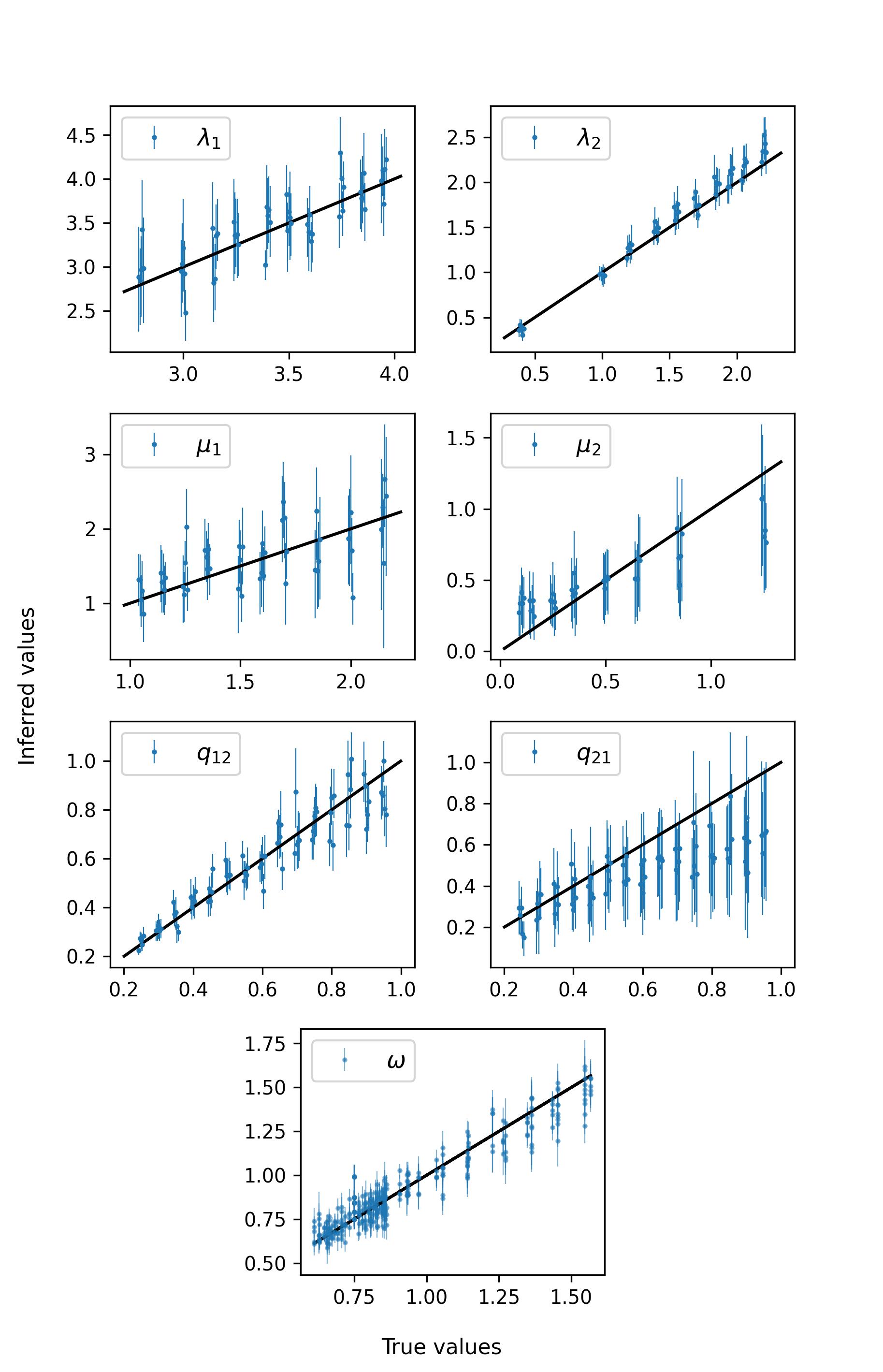}
	\caption{Inferred vs true parameter values for the irreducible case. One parameter is changed at a time, while the other parameters stay fixed at their default values. Error bars represent the 50\% credible intervals for the final posteriors.}
	\label{fig:6pparams}
\end{figure}

Figure~\ref{fig:6pparams} shows our results when each parameter is varied one at a time, with 5 replicates for each value. As with the general reducible case, the estimation of the phase-1 birth rate becomes more accurate as the rate becomes larger, while the opposite is true for the phase-2 birth rate. Again, there is a small positive bias for the phase-2 birth rate. The inference of both phase-1 and phase-2 death rates becomes less accurate as the rates increase; in addition, the phase-2 death rate is overestimated for small values and underestimated for larger values. Finally, as the transition rates increase, their accuracy decreases, and there is an underestimation of the phase 2 to phase 1 transition rate as it becomes large. Like most parameters, the accuracy of the growth rate decreases as the value increases, but there is no noticeable bias.

The general patterns observed here are mostly consistent with the reducible cases. 
While we can still estimate the rates satisfactorily, we note that the performance in this irreducible case is slightly less optimal compared to the simpler reducible cases, which is expected. Nonetheless, our method shows great potential in dealing with more complex scenarios.

We also studied the sensitivity of the inference for different tree sizes (number of leaves). In Supplementary Figure~\hyperlink{text:size}{S17}, we show the parameter inference for different tree sizes (with 25 trees in total) in the observed dataset. As expected, as the size of the trees increases (thus increasing the amount of information in the dataset), the inference becomes more accurate and less biased. This effect is less apparent, but still present, for the phase-1 rates.

In Supplementary Figure~\hyperlink{text:size2}{S18}, we vary tree size while keeping the total number of leaves in all observed trees fixed at 5000 (thus nominally keeping the amount of information the same in the datasets). There is relatively little change compared to the previous case, although there appears to be marginally better estimation for datasets with more trees and a smaller number of leaves.

\subsection{Comparison with maximum likelihood methods}

We compare the performance of our ABC method with the maximum likelihood estimates (MLEs) from the BiSSE model for the six-parameter irreducible case, as implemented in the \texttt{castor} package \cite{louca2020general}.

\begin{figure}[t]
    \centering
    \includegraphics[height=5cm]{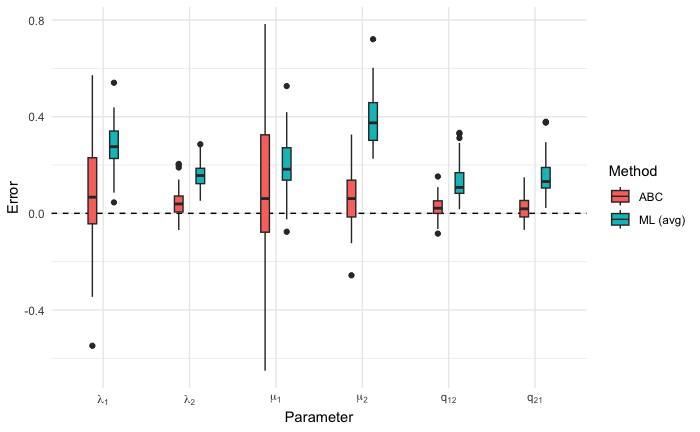}
    \caption{Parameter inference error for ABC and ML methods for the irreducible case, for the default parameters for $50$ replicates.}
    \label{bp:e_n100_d_MLavg}
\end{figure}

We first consider the default parameters $(\lambda_1,\lambda_2,\mu_1,\mu_2,q_{12},q_{21}) = (3,1,2,0.5,0.5,0.25)$ as in Section \ref{sec:irred}. As before, our observed dataset consists of 100 trees, each with 50 leaves. Because the observed dataset has multiple trees, but \texttt{castor} calculates the MLEs for a single tree, we average the MLEs for each parameter over the 100 trees to generate the final estimate. The results are shown in Figure \ref{bp:e_n100_d_MLavg} for 50 replicates. In all cases, it is clear that the ABC estimates are less biased than the MLEs.
In Table \ref{tab:rrmse}, we show the relative root mean squared error (RRMSE) of the estimates. For all parameters except $\mu_1$, the ABC method outperforms the maximum likelihood method; the performance for $\mu_1$ appears to be the result of a larger variance in the ABC estimates overcoming a smaller bias.

\begin{table}[htb]
 \begin{tabular}{|l|c|c|c|c|c|c|c|} 
 \hline
 RRMSE & $\lambda_1$ & $\lambda_1$ & $\mu_1$ & $\mu_2$ & $q_{12}$ & $q_{21}$\\ [0.5ex] 
 \hline\hline
ABC& 0.08 & 0.08 & 0.16 & 0.26 &  0.10 &  0.20\\  
ML & 0.10& 0.17 & 0.12 & 0.79 & 0.30&  0.69\\ [1ex]
 \hline
 \end{tabular}
 \caption{RRMSE for ABC and ML methods for the irreducible case, for the default parameters for $50$ replicates.}\label{tab:rrmse}
\end{table}

To compare the performance of the methods over a wider range of parameters, we also varied the true parameters so that they were simultaneously drawn from the prior distributions Unif$(0,5)$, under the constraint that the MBT be supercritical. We then performed inference as before, for datasets of 100 trees with 50 leaves each.

\begin{figure}
\hspace{-2cm}
\begin{tabular}{cc}
  \includegraphics[width=75mm]{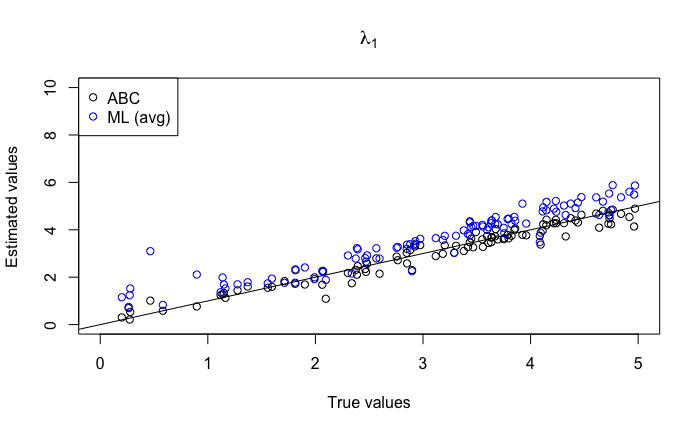} &   \includegraphics[width=75mm]{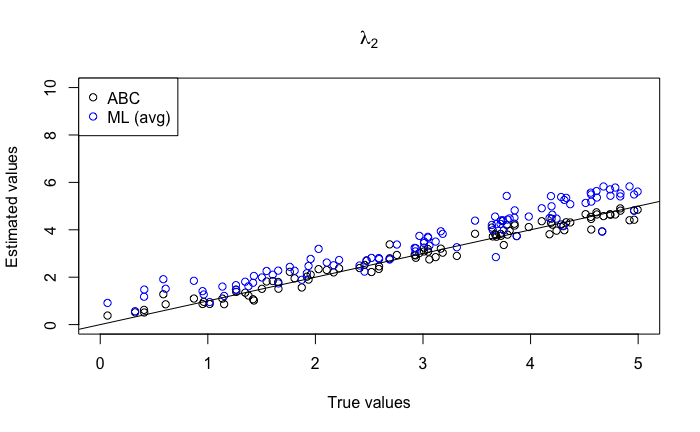}\\
 \includegraphics[width=75mm]{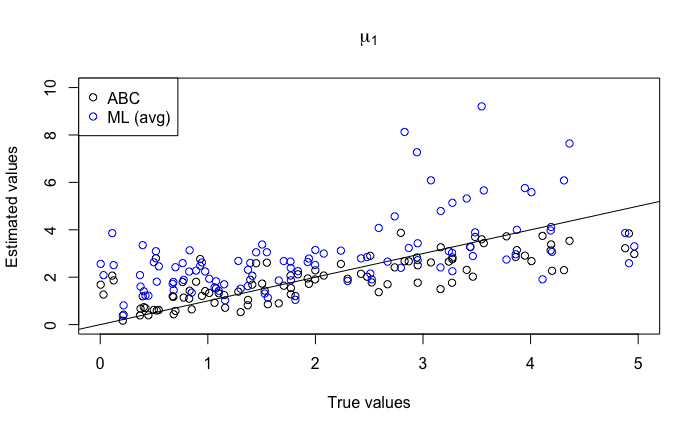} &   \includegraphics[width=75mm]{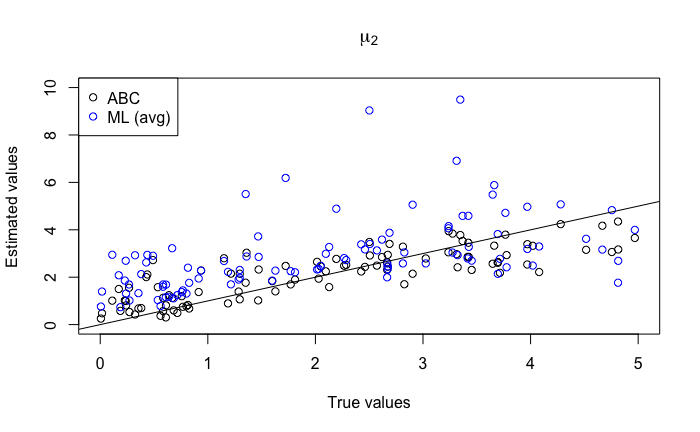} \\
 \includegraphics[width=75mm]{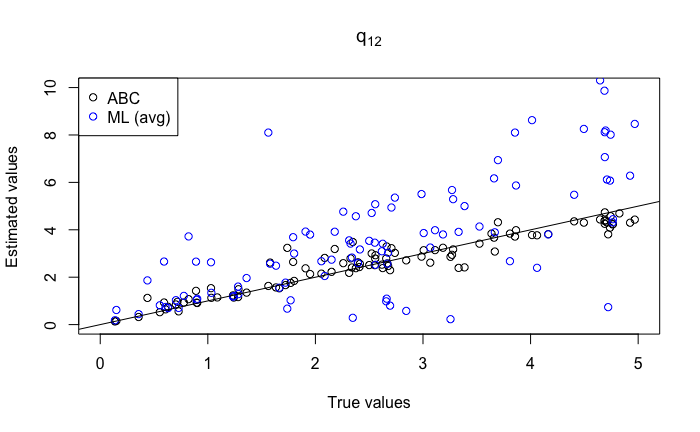} &   \includegraphics[width=75mm]{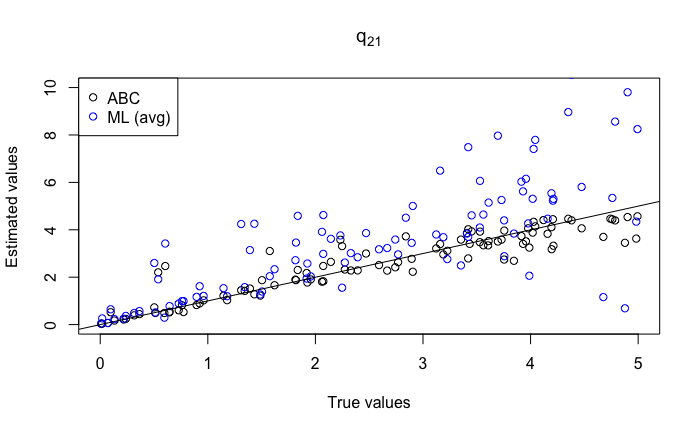} 
\end{tabular}
\caption{Inferred vs true parameter values for ABC and ML methods for the irreducible case, with varying parameters for $100$ replicates. Seven samples with at least one inferred ML transition rate exceeding 10 have been excluded from the plot.}
\label{figs:n100_diff_compare}
\end{figure}

The results for 100 replicates are shown in Figure \ref{figs:n100_diff_compare}. As before, we can observe a distinct bias for the MLEs; the birth rates $\lambda_1$ and $\lambda_2$ are inferred with positive bias, as are the transition rates $q_{12}$ and $q_{21}$. Additionally, the variance of the inferred transition rates increases as the parameter values increase. The numerical approximation of MLEs can give poor estimates and lead to unreasonably high transition rates (some transition rates were estimated to be greater than 100; these outliers are not shown in the figure). In contrast, the ABC estimates are significantly more precise and show no discernible bias. For the death rates $\mu_1$ and $\mu_2$, there appears to be a positive bias at low parameter values and a negative bias at high parameter values for both methods, as was observed for the ABC method in Section \ref{sec:irred}. However, this effect is again much less noticeable in the ABC estimates. In Table \ref{tab:rrmse2}, we show the RRMSE for these estimates. In all cases, it appears that the ABC method substantially outperforms the ML method, particularly for the transition rates.

\begin{table}[htb]
 \begin{tabular}{|l|c|c|c|c|c|c|c|} 
 \hline
 RRMSE & $\lambda_1$ & $\lambda_1$ & $\mu_1$ & $\mu_2$ & $q_{12}$ & $q_{21}$\\ [0.5ex] 
 \hline\hline
ABC & 0.10 & 0.09 & 0.41 & 0.38  & 0.16   & 0.21\\ 
ML & 0.21 & 0.22 & 0.79 & 0.79 &  0.90 &  0.91\\ [1ex]
 \hline
 \end{tabular}
 \caption{RRMSE for ABC and ML methods for the irreducible case, with varying parameters for $100$ replicates. Three samples with unreasonable ML transition rates ($>100$) have been excluded.}\label{tab:rrmse2}
\end{table}

To verify that similar results apply to problem sizes closer to those in the real data analysis in the following section, we repeated this analysis using observed (and simulated) datasets consisting of a single tree with 5000 leaves. The results are shown in Supplementary Figures \hyperlink{text:e_large_d_comp}{S19}--\hyperlink{text:large_diff_compare}{S20}, and the RRMSE is shown in Supplementary Tables \hyperlink{table:rrmse_default_s5000}{S1}-\hyperlink{table:rrmse_diff_s5000}{S2}. The results are consistent with our previous experiments, showing ABC estimates that have higher accuracy than ML estimates.


\section{Real data analysis}\label{sec:realdata}

We apply our method to a published time-calibrated 3951-tip phylogenetic tree of squamata (reptiles) from \cite{pyron2013phylogeny}, shown in Figures 2--28 of that paper. The species are classified according to their method of bearing offspring, with 3108 oviparous (egg-laying) and 843 viviparous (live-bearing) species. We assign oviparity to phase 1 and viviparity to phase 2. This classification is of particular interest as (a) it allows us to unambiguously assign biologically meaningful phases, and (b) there is a clear direction of evolution, with oviparity known to be the ancestral state and viviparity evolving from oviparity. However, it is not obvious if it is possible to revert back from viviparity to oviparity; this suggests that either a reducible or irreducible model may be used here.

The squamata dataset was analysed in \cite{pyron2014early} using the BiSSE model. The authors estimated the diversification and transition rates using the maximum likelihood method, and found evidence for an early transition to viviparity at a basal lineage, together with multiple reversions from viviparity to oviparity. They compared multiple models, including a reducible model disallowing a viviparity-to-oviparity transition, and concluded that an irreducible model allowing transitions in both directions fitted the data best. 

We investigate the diversification and transition rates under an irreducible model with our ABC-PMC method. In this case, the observed data consist of a single large tree instead of multiple smaller trees with the same tree size; as a result, a few adjustments have been made to the algorithm (details are provided in Supplementary Section \ref{sec:onetree}). Based on the MLEs in \citet{pyron2014early}, we used Unif$(0,0.2)$ as the prior for all parameters. Since it has been shown that the most recent common ancestor of the species is viviparous (phase 2) with strong support \citep{pyron2014early}, we use this as the starting phase. 

\begin{figure}\centering
	\includegraphics[width=\textwidth]{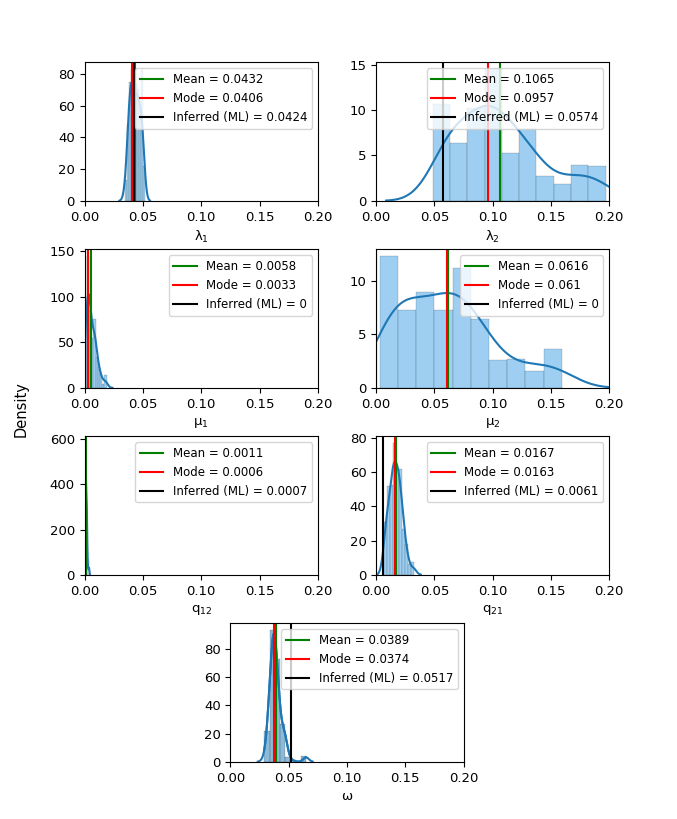}
	\caption{Final posterior distributions for the parameters and growth rate of the squamata phylogeny, with mean (green), mode (red), and the inferred values using the Maximum Likelihood (ML) method (black), assuming the process starts in phase $2$ and all extant species have been sampled.}
		\label{fig:posterior_parity}
		
\end{figure}

In Supplementary Figure \hyperlink{text:paritytrace}{S21}, we present the trace plots of the posterior means for the parameters and the growth rate. These plots indicate convergence to a stable posterior by the 30th iteration. In Figure~\ref{fig:posterior_parity}, we show the final posterior distributions for the parameters, together with the MLEs as calculated by \texttt{castor}. Note that these differ from those calculated in \cite{pyron2014early} as we use a sampling fraction of 1 (i.e., all extant species have been sampled) to be consistent with our ABC method. The diversification rates in phase $2$ are estimated less accurately than those in phase $1$, as we have fewer phase-2 leaves. We find that viviparous species have higher estimated speciation and extinction rates than oviparous species, consistent with the findings of \cite{pyron2014early}. 
This suggests that our method is able to capture the signals of rate variation in the data. We further show that most of the observed summary statistics stay well within the range of the posterior predictive distributions, indicating a reasonable fit; see Supplementary Section \ref{sec:gof}.

\begin{figure}
      \centering
	\includegraphics[width=0.6\textwidth]{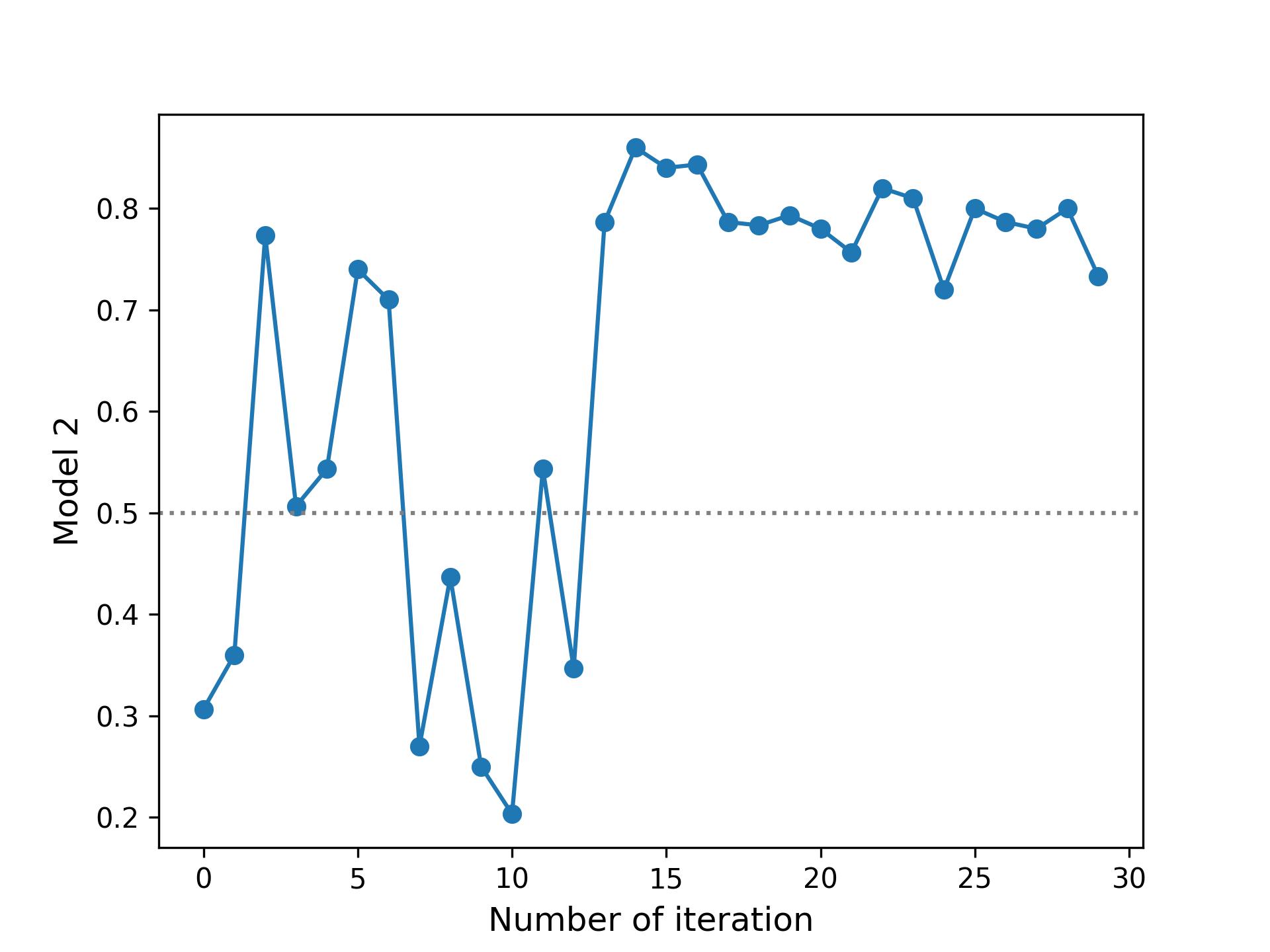}
	\caption{The proportion of accepted samples from model 2 (the irreducible model) against the reducible model at each iteration.}
		\label{fig:mod_sel}
\end{figure}

We also apply ABC Sequential Monte Carlo (ABC SMC) model selection \citep{toni2009approximate} to compare the irreducible model to a reducible model where the viviparity-to-oviparity transition is prohibited. For this purpose, we increase the number of accepted samples to $N=300$ at each iteration to ensure a large enough number of samples. In Figure \ref{fig:mod_sel}, we see that the irreducible model is chosen over the reducible model with a final posterior probability of 0.73. This again supports the conclusion of \citet{pyron2014early}, who found evidence for transitions in both directions.


\section{Discussion}

In this paper, we used an ABC scheme to infer the diversification rates in Markovian binary tree models from phylogenetic trees. We developed appropriate summary statistics that enable us to capture variation in the birth and death rates in each phase, and in the transition rates between phases. In the cases we studied, limited to two phases, we found that we can generally infer the true parameters with high accuracy, particularly in the simpler reducible case. This suggests that MBTs can be practically used as a probabilistic model for phylogenetic trees, and that ABC methods have great potential for the inference of MBT parameters in applied contexts.

We also compared the performance of our ABC method to the existing maximum likelihood (ML) method for the equivalent BiSSE model. In general, our ABC method outperforms the ML method in terms of relative root mean squared error. 
This could be attributed to the fact that ML methods provide only a single point estimate without accounting for the broader likelihood `landscape'. In contrast, a Bayesian approach, even when estimating a posterior distribution approximately, considers the full range of parameter uncertainty, leading to a more accurate overall estimate.
Additionally, the use of heuristic methods to maximise the likelihood can also impact the performance of the MLE. In terms of computational cost, ABC has higher computational time than the ML approach, as it relies on simulations. However, the simulation process can fully parallelised, reducing the computational cost significantly. 

Several issues remain to be explored. For both the `real' observations and the simulated samples, simulated trees are discarded if they go extinct; this may introduce bias in both the `true' parameter values and their inference, although our results suggest that this effect is small. Likewise, any simulated parameter values are discarded if the corresponding overall growth rate $\omega$ is less than 0 (resulting in a subcritical process). Again, this may cause some bias in the true prior, which needs further investigation.

The scope of the MBT models used in this paper is somewhat limited in nature, as they are restricted to just two phases and do not allow all possible transitions; an unrestricted 2-phase model requires 12 free parameters. We have limited ourselves to smaller models to first establish that accurate inference is possible in these cases, and our results indicate that this is indeed the case. The restricted version of the MBT model used here is equivalent to the 2-state BiSSE model, but the full model is far more flexible. 

There is no theoretical barrier to extending the ABC method to an MBT model with a larger number of phases, or an unrestricted 2-phase model. However, our results suggest that inference in these larger cases may become challenging (not unlike the BiSSE series of models). A potential limiting factor is that the ABC method requires the number of summary statistics to be at least as large as the number of parameters being inferred.  For these cases, developing additional appropriate summary statistics may be necessary to achieve accurate inferences.


Lastly, our method can be further applied to other real datasets. The squamata phylogeny itself was re-analysed in \cite{halliwell2017live}, with phases representing both ovi-/viviparity and social grouping. In theory, our model could be extended here to include 4 phases (all possible combinations), although the corresponding impact on accuracy is unclear. 
For other datasets, phases may not need to correspond to any particular phenotype, but could simply provide extra degrees of flexibility in the modelling. This also suggests that we may need to extend our method to account for cases where phases are not observed in extant species.

\section*{Competing interests}
No competing interest is declared.

\section*{Acknowledgements}

Sophie Hautphenne would like to thank the Australian Research Council (ARC) for support through her Discovery Project DP200101281.

\bibliographystyle{abbrvnat}
\bibliography{mbt.bib}

\clearpage
\newpage
\section*{\centering Supplementary material for ``Approximate Bayesian computation for Markovian binary trees in phylogenetics"}
\beginsupplement

\section{Supplementary Methods}

\subsection{Markovian binary trees}\label{sec:mbt}

Markovian binary trees (MBTs) form a flexible class of continuous-time Markovian multitype branching processes \citep{athreya1972multi} in which the lifetime and reproduction process of each individual in the population is controlled by an underlying transient Markov chain with $n$ transient states, called \textit{phases}, and one absorbing state 0. While in phase $i\in\{1,\dots,n\}$, an individual can:
\begin{itemize}
	\item transition to phase $j\neq i$ at constant rate $(D_0)_{ij}$;
	\item give birth to a child in phase $k$ and simultaneously transition to phase $j$ at constant rate $B_{i,kj}$;
	\item die, that is, enter the absorbing state 0, at constant rate $d_i$.
\end{itemize}
Thus an MBT is parameterised by an $n \times n$ matrix $D_0$, an $n \times n^2$ matrix $B$, and an $n \times 1$ vector $\d$. The diagonal elements of $D_0$ are negative and are such that
\[D_0 \1 + B \1 + \d = \0,\]
where $\1$ and $\0$ are vectors of 1's and 0's, respectively, of the appropriate size. 

MBTs are `matrix' generalisations of birth-and-death processes: in an MBT, the lifetime of individuals is distributed according to a \textit{phase-type} distribution, which is a generalisation of the exponential distribution, and the reproduction process of the individuals follows a (transient) \textit{Markovian arrival process}, which is a generalisation of the Poisson process (see \citealp{latouche1999introduction}).
Note that while the MBT literature usually considers each lineage as an individual organism, in this paper they represent species, where birth events represent speciations and death events represent extinctions. 

It can be shown \citep{hautphenne2009transient} that the mean number of individuals in phase $j$ in the population at time $t$, given that the population started with a single individual in phase $i$ at time 0, is given by the $(i,j)$th entry of the mean population size matrix  
\[M(t)= \exp(At),\]
where 
\[A = D_0 + B(\1 \otimes I + I \otimes \1),\]
and $\otimes$ denotes the Kronecker product. The asymptotic behaviour of the MBT therefore depends on the dominant eigenvalue of $A$, which we denote by $\omega$ and call the \emph{growth rate}. There are three cases:
\begin{itemize}
	\item when $\omega < 0$, the mean population size goes to 0 as $t \rightarrow \infty$, and the population eventually becomes extinct with probability 1 (\emph{subcritical});
	\item when $\omega = 0$, the asymptotic mean population size is constant (\emph{critical}), and the population eventually becomes extinct with probability 1;
	\item when $\omega > 0$, the mean population size grows without bounds, and the population has a positive probability of ultimate survival (\emph{supercritical}).
\end{itemize}
We are primarily interested in the supercritical case, which is suitable to represent the exponential increase in the variety of species observed in real life.

In this paper, we concentrate on the special case where $n = 2$, that is, there are only two possible phases. We further assume that individuals stay in the same phase when giving birth, and give birth to children in that same phase. Thus the matrices $D_0$, $B$, and $\d$ are given by
\[D_0 = \begin{pmatrix} * & q_{12} \\ q_{21} & * \end{pmatrix}, \, B = \begin{pmatrix} \lambda_1 & 0 & 0 & 0 \\ 0 & 0 & 0 & \lambda_2 \end{pmatrix}, \, \d = \begin{pmatrix} \mu_1 \\ \mu_2 \end{pmatrix},\]
	where $\lambda_1$ and $\lambda_2$ are the birth rates for each phase, $\mu_1$ and $\mu_2$ are the death rates for each phase, and $q_{12}$ and $q_{21}$ are the transition rates between phases. We denote this set of parameters by $\theta = (\lambda_1, \lambda_2, \mu_1, \mu_2, q_{12}, q_{21})$.
We assume $q_{12}>0$. If individuals in phase 2 cannot transition to phase 1 (i.e., $q_{21} = 0$), we call the process \emph{reducible}; otherwise, we call it \emph{irreducible}.

\subsection{ABC-PMC}\label{sec:abc-pmc}

The ABC-PMC method \citep{beaumont2009adaptive} is an extension of the basic ABC rejection method \citep{beaumont2002approximate}. We use as input to the method a set of $m$ `real' observed phylogenetic trees $Y_{\text{obs}}$, with branch lengths given in units of substitutions per site (typically these would be inferred, potentially with some error, from sequence data). We assume that these trees are ultrametric and that the phases of the leaves are known (but the phases are unobserved in other parts of the tree).

The algorithm proceeds over a series of $T$ iterations. In the first iteration $t = 0$, a set of parameters $\theta^*$ are proposed from a prior distribution $\boldsymbol\pi(\theta)$. In further iterations ($t \geq 1$), we adapt the posterior samples from the previous iteration to construct a new prior distribution as detailed in Algorithm \ref{alg:cap}, and then propose a set of parameters $\theta^*$ from this prior.

At iteration $t$, we simulate $n_t$ phylogenetic trees using an MBT process with parameters $\theta^*$, and calculate a set of summary statistics (detailed in Section \ref{sec:summary} of the manuscript) for these trees. Given a set of tolerance values $\boldsymbol\epsilon_t=\{\epsilon_{tk}\}_{k=1}^K$, where $\epsilon_{tk}$ is the tolerance for the $k$th summary statistic, and $K$ is the total number of summary statistics, we accept the proposed parameters if the absolute difference between each summary statistic for the simulated and observed trees is less than the corresponding tolerance value; 
otherwise, we reject them. We continue until we have accepted $N$ samples; this results in a `population' of accepted parameters  $\{\theta^{(i)}_t\}_{i=1}^N$, which provides an approximation to the posterior distribution of the parameters.

To propose parameters for iteration $t\geq1$, we draw one of the $N$ accepted parameter sets from the previous iteration, with weights $\{\omega^{(i)}_{t-1}\}_{i = 1, \ldots, N}$. These weights are calculated at the end of iteration $t-1$  (details on how to calculate them at the end of the current iteration $t$ are provided below). We then perturb the sample by adding a normal perturbation $\mathcal{N}(0, \Sigma_{t-1})$, where $\Sigma_{t-1}$ is defined below, to produce the proposed parameters.

At the end of the iteration, we calculate the weights for the next iteration as follows. If $t = 0$, we take $\omega_0^{(i)} = \frac1N$ for all $i\in\{1,2,\ldots,N\}$, so that all samples are weighted equally. If $t\geq 1$, we let 
\[\omega_t^{(i)} \propto
\frac{ \boldsymbol\pi(\theta^{(i)}_t)  }{ \sum_{j=1}^N \omega_{t-1}^{(j)} f(\theta_t^{(i)};\theta_{t-1}^{(j)},\Sigma_{t-1}) },\, \qquad i\in\{1,2,\ldots,N\},\]
where $f(\,\cdot\,;{\mu},\Sigma)$ is the multivariate Gaussian density function, and $\Sigma_{t-1}$ is twice the weighted covariance matrix of the posterior samples from iteration $t-1$:
\begin{align*}
    \Sigma_{t-1} &= 2\, \text{var}\left\{ \theta^{(i)}_{t-1}, \omega^{(i)}_{t-1}\right\} \\
&= 2 \sum_{i=1}^N \omega_{t-1}^{(i)}\left[\theta^{(i)}_{t-1}-\left(\sum_{i=1}^N\omega_{t-1}^{(i)}\theta^{(i)}_{t-1}\right)\right]
\left[\theta^{(i)}_{t-1}-\left(\sum_{i=1}^N\omega_{t-1}^{(i)}\theta^{(i)}_{t-1}\right)\right]^\top.
\end{align*}
The weights $\{\omega^{(i)}_t\}_{i = 1, \ldots, N}$ are then normalised so that their sum is 1.

The full formal details of the algorithm are given in Algorithm \ref{alg:cap}.

\begin{algorithm}
\caption{ABC-PMC sampling}\label{alg:cap}
\begin{algorithmic}

\State \makebox[0.9cm][l]{$\mathtt{S1}$} Initialize  $\epsilon_{1k} \geq \epsilon_{2k} \geq \cdots \geq \epsilon_{Tk}, \forall k \in \{1,\cdots,K\}$.
\State \makebox[0.9cm][l]{} Set the iteration indicator $t=0$.
\State \makebox[0.9cm][l]{$\mathtt{S2.0}$} Set the sample indicator $i=1$.
\State \makebox[0.9cm][l]{$\mathtt{S2.1}$} If $t=0$, draw a sample $\theta^{**}$ from the prior $\boldsymbol\pi(\theta)$
\State \makebox[0.9cm][l]{} If $t\neq0$, draw a sample $\theta^{*}$ from the previous population $\{\theta_{t-1}^{(i)}\}_{i=1}^N$ with weights $\omega_{t-1}$ and perturb this sample using a normal distribution, $\theta^{**} \sim \mathcal{N}(\theta^*, \Sigma_{t-1})$.
\State \makebox[0.9cm][l]{$\mathtt{S2.2}$} If $\boldsymbol\pi(\theta^{**}) = 0$, where  $\boldsymbol\pi(\cdot)$ is the density function of the prior distribution, return to step $\mathtt{S2.1}$.
\State \makebox[0.9cm][l]{$\mathtt{S2.3}$} Generate a dataset $Y^{(i)}$ from the model with parameters $\theta^{**}$.
\State \makebox[0.9cm][l]{$\mathtt{S2.4}$} If $|s_k(Y_{\text{obs}}) - s_k(Y^{(i)})| > \epsilon_{tk}$ for some $k$, where $s_k(\cdot)$ represents the $k$th summary statistic, return to step $\mathtt{S2.1}$.
\State \makebox[0.9cm][l]{$\mathtt{S2.5}$} Set $\theta_t^{(i)} = \theta^{**}$, and compute the weight $\omega_{t}^{(i)}$ for the parameter $\theta_t^{(i)}$:\\
\makebox[0.32cm][l]{} If $t=0$, $\omega_{t}^{(i)} \propto 1$.\\
\makebox[0.35cm][l]{} If $t\neq0$, $\omega_{t}^{(i)} \propto \boldsymbol\pi(\theta_t^{(i)})/\sum_{j=1}^N\omega_{t-1}^{(j)}\,f(\theta_t^{(i)};\theta_{t-1}^{(j)},\Sigma_{t-1})$.
\State \makebox[0.9cm][l]{$\mathtt{S2.6}$} If $i < N$, set $i = i+1$. Return to step $\mathtt{S2.1}$.
\State \makebox[0.9cm][l]{$\mathtt{S3.0}$} Normalise the weights.
\State \makebox[0.9cm][l]{$\mathtt{S3.1}$} Set $\Sigma_{t}=2\,\text{var}\{\theta_{t}^{(i)},\omega_{t}^{(i)}\}_{i=1}^N$.
\State \makebox[0.9cm][l]{$\mathtt{S3.2}$} If $t < T$, set $t = t+1$. Return to step $\mathtt{S2.0}$.

\end{algorithmic}
\end{algorithm}

\subsection{Method details for simulation study}

When we apply the ABC-PMC method to simulated data, we use the following values:
\begin{itemize}
    \item $T = 30$ total iterations;
    \item prior distribution Unif(0,5) for all parameters.
\end{itemize}

The initial tolerance values $\boldsymbol{\epsilon}_0$ are estimated from the real trees $Y_{\text{obs}}$. For each summary statistic, the corresponding tolerance value is calculated separately; here, we set them to be twice the standard deviation of the summary statistic across the trees in the observed dataset. To maintain a reasonable acceptance rate, subsequent tolerance values are a constant multiple (here we take $e^{-0.2}$)  of the preceding values if the acceptance rate is above a specified threshold (we take $0.03$); otherwise, they are left unchanged. 

For irreducible processes, considering the increase in the number of parameters, we use a wider acceptance region in the first iteration, which allows us to obtain the required number of accepted samples in a reasonable time. In the first iteration, if the summary statistics of the proposed sample stays within the region spanned by the minimal and maximal values of statistics from the observed trees, it will be accepted. 

When proposing parameters, it is possible that the proposed parameters generate a critical or subcritical process, i.e., the growth rate $\omega \leq 0$. If this happens, we automatically reject the proposed parameters.

Likewise, it is possible that the proposed parameters generate a supercritical process, but one that is only `slightly' so, i.e., has a high probability of going extinct. If we generate a tree that becomes entirely extinct before it reaches the desired number of leaves (50 in the default case), we discard it and generate another tree. If this occurs for the first 25 times in a row for a set of proposed parameters, we reject those parameters.

\section{Sensitivity analysis}\label{sec:sensitivity}

\begin{figure}
    \centering
    \includegraphics[width=1\linewidth]{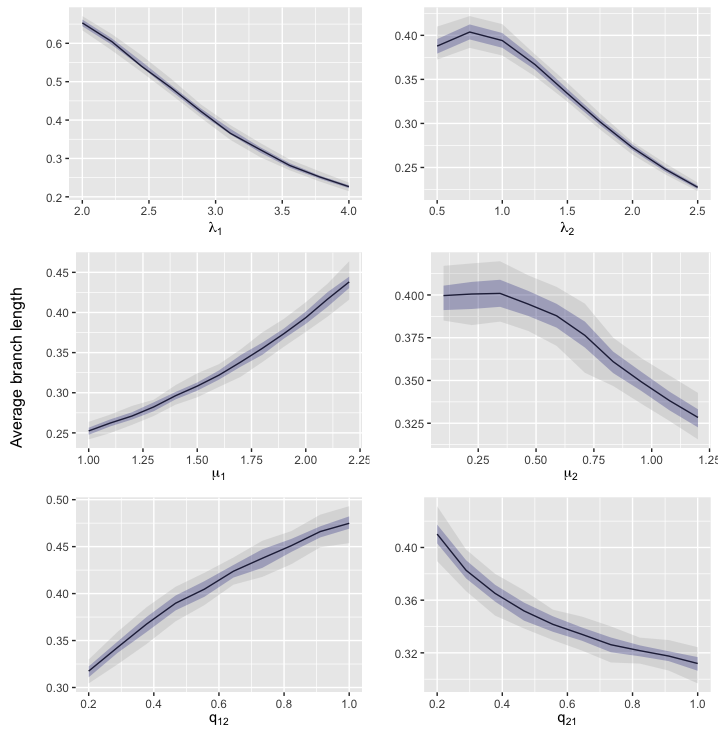}
    \caption{Sensitivity of average branch length. The solid lines indicate the mean, blue regions bound the $0.25$th and $0.75$th quantiles, while grey regions bound the $0.05$th and $0.95$th quantiles.}
    \label{fig:sensit_stat5}
\end{figure}

\begin{figure}
    \centering
    \includegraphics[width=1\linewidth]{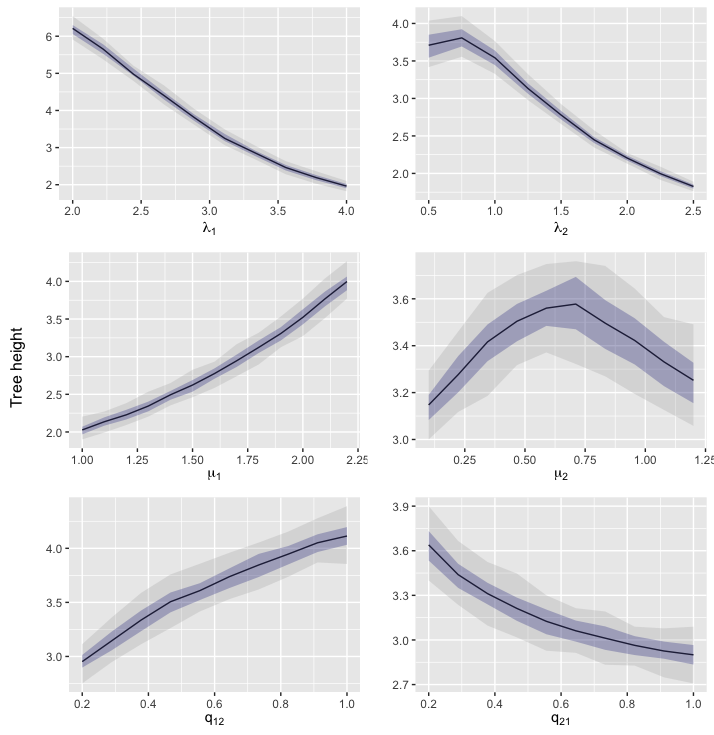}
    \caption{Sensitivity of tree height. The solid lines indicate the mean, blue regions bound the $0.25$th and $0.75$th quantiles, while grey regions bound the $0.05$th and $0.95$th quantiles.}
    \label{fig:sensit_stat4}
\end{figure}

\begin{figure}
    \centering
    \includegraphics[width=1\linewidth]{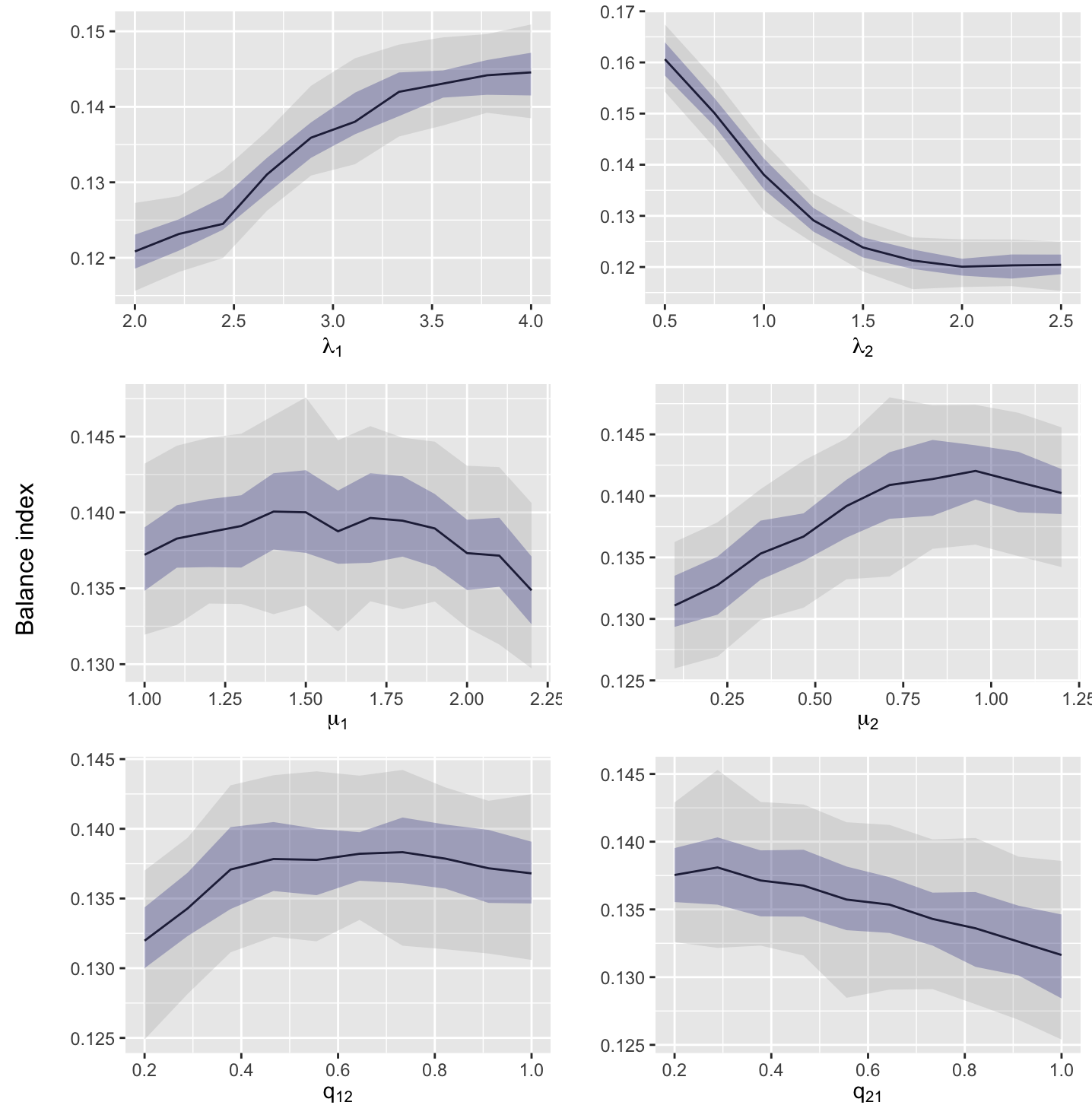}
    \caption{Sensitivity of balance index. The solid lines indicate the mean, blue regions bound the $0.25$th and $0.75$th quantiles, while grey regions bound the $0.05$th and $0.95$th quantiles.}
    \label{fig:sensit_stat1}
\end{figure}

\begin{figure}
    \centering
    \includegraphics[width=1\linewidth]{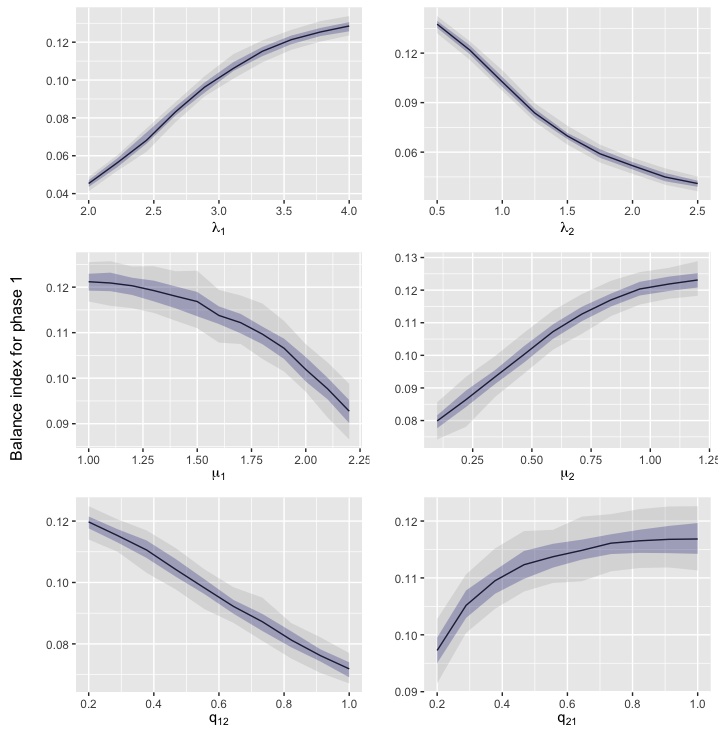}
    \caption{Sensitivity of balance index for phase 1. The solid lines indicate the mean, blue regions bound the $0.25$th and $0.75$th quantiles, while grey regions bound the $0.05$th and $0.95$th quantiles.}
    \label{fig:sensit_stat2}
\end{figure}

\begin{figure}
    \centering
    \includegraphics[width=1\linewidth]{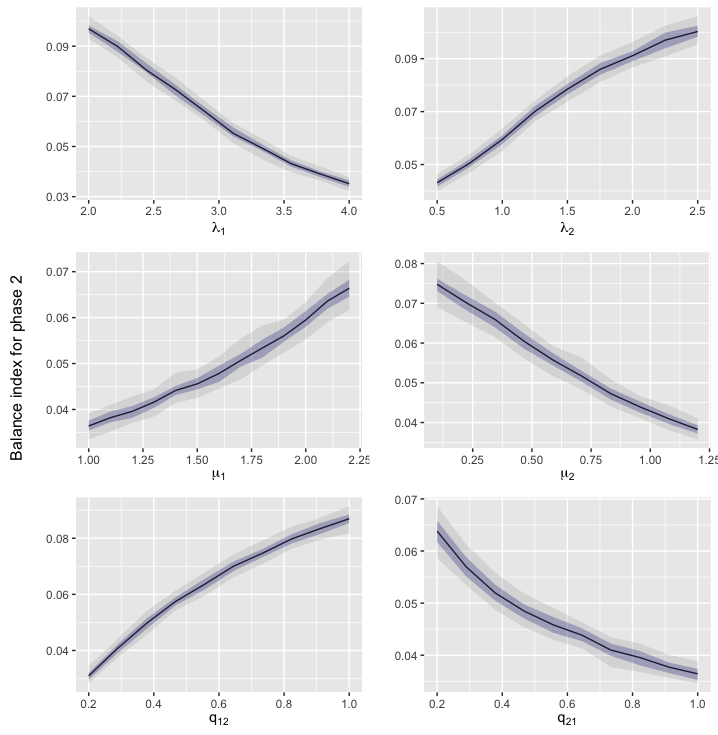}
    \caption{Sensitivity of balance index for phase 2. The solid lines indicate the mean, blue regions bound the $0.25$th and $0.75$th quantiles, while grey regions bound the $0.05$th and $0.95$th quantiles.}
    \label{fig:sensit_stat3}
\end{figure}

\begin{figure}
    \centering
    \includegraphics[width=1\linewidth]{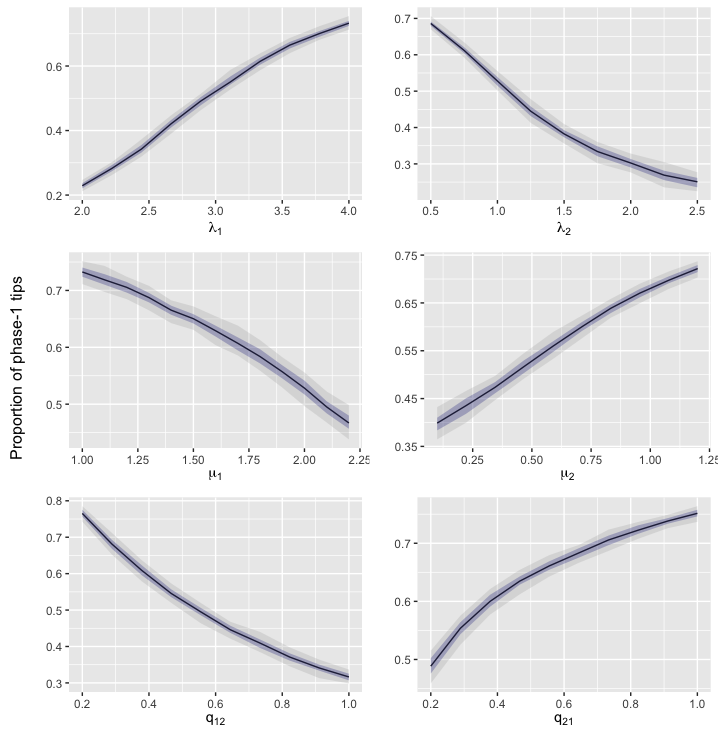}
    \caption{Sensitivity of proportion of phase-1 tips. The solid lines indicate the mean, blue regions bound the $0.25$th and $0.75$th quantiles, while grey regions bound the $0.05$th and $0.95$th quantiles.}
    \label{fig:sensit_stat6}
\end{figure}

\begin{figure}
    \centering
    \includegraphics[width=1\linewidth]{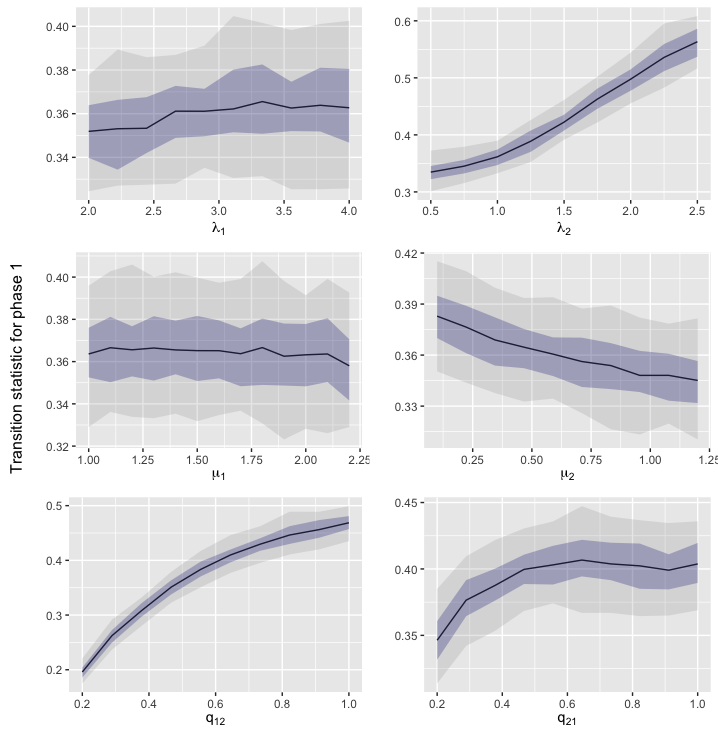}
    \caption{Sensitivity of transition statistic for phase 1. The solid lines indicate the mean, blue regions bound the $0.25$th and $0.75$th quantiles, while grey regions bound the $0.05$th and $0.95$th quantiles.}
    \label{fig:sensit_stat7}
\end{figure}

\begin{figure}
    \centering
    \includegraphics[width=1\linewidth]{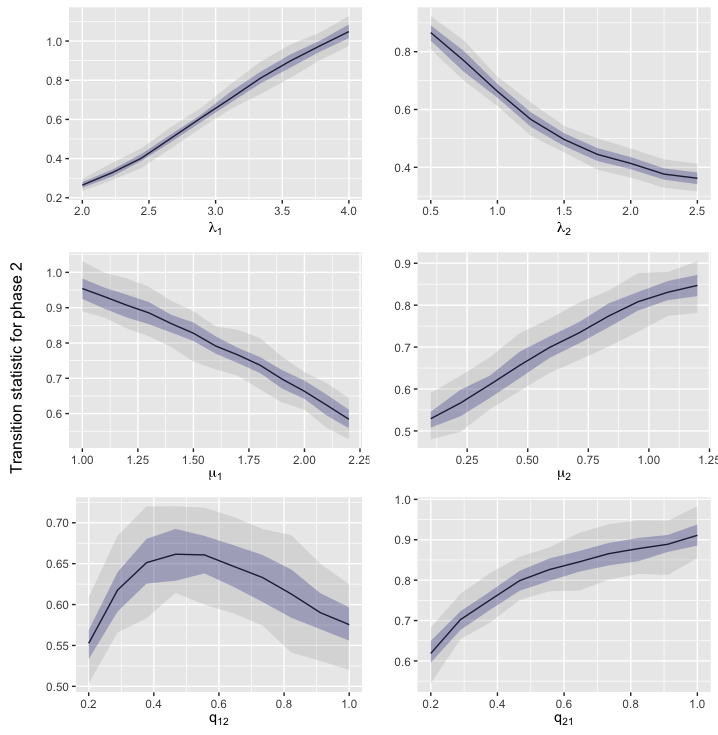}
    \caption{Sensitivity of transition statistic for phase 2. The solid lines indicate the mean, blue regions bound the $0.25$th and $0.75$th quantiles, while grey regions bound the $0.05$th and $0.95$th quantiles.}
    \label{fig:sensit_stat8}
\end{figure}

Here we investigate the sensitivity of the selected summary statistics to changes in the parameter values. We use the default parameter values as the baseline, $(\lambda_1, \lambda_2, \mu_1, \mu_2, q_{12}, q_{21})=(3,1,2,0.5,0.5,0.25)$, and change one parameter at a time. The range of the parameter values are taken from the prior, under the condition that the MBT process is supercritical. For each parameter set, we generate 100 datasets (each containing 100 trees with 50 leaves each) and estimate the quantiles of the distribution for each summary statistic. The results are shown in Figures \ref{fig:sensit_stat5}--\ref{fig:sensit_stat8}. Here, we see that each of the selected summary statistics are sensitive to at least one of the parameters, with most being sensitive to all of them.

\section{Inference from one observed tree}\label{sec:onetree}

For the analysis of the squamate phylogeny (Section \ref{sec:realdata} of the manuscript), we make some changes to the method to account for the fact that there is only one observed tree.

\paragraph{Tolerance values}\label{par:tolone}

With a single observed tree, the initial tolerance values can no longer be determined by the variance of summary statistics in the observed dataset. 
\begin{itemize}
    \item For the first iteration, the tolerance level is constructed based on the observed data. We break the large observed tree (with $\approx 4000$ tips) into several small subtrees with 20--50 leaves each, as illustrated in \autoref{fig:breaktree}, and compute the corresponding summary statistics for each subtree. 
    To explore the parameter space efficiently, we use a simplified simulation process to approximate the underlying model in order to reduce the computational cost of generating datasets for each proposed sample.
    We select an arbitrary subtree among the observed subtrees.
    For each set of proposed parameters, $\boldsymbol{\theta}^{(i)}$, we grow a tree that has the same size as the selected subtree. If the summary statistics of the simulated tree stays within the range spanned by the minimal and maximal values of the statistics in the observed subtrees, the sample will be accepted. 
    
    \item From the second iteration, we set the initial tolerance vector, $\boldsymbol\epsilon_0$, to be two times the standard deviation of the corresponding summary statistics in the observed subtrees. 
    This allows us to accept the required number of samples in a reasonable timeframe. 

    \begin{figure}
	\centering
	\includegraphics[width=0.75\textwidth]{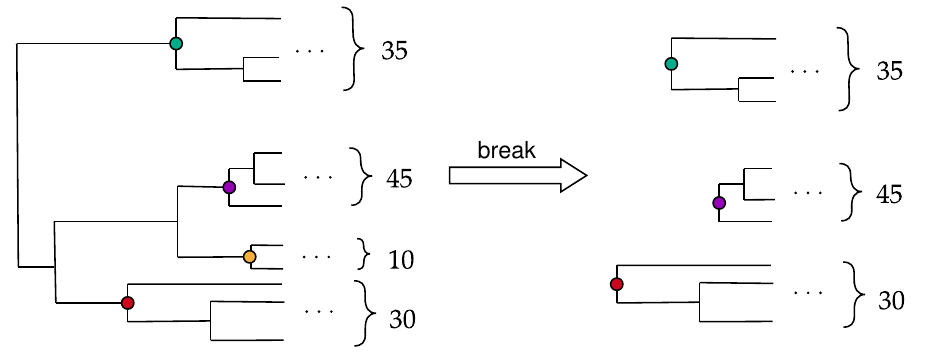}
 \caption{An example of breaking one large tree into several small trees. Numbers indicate the number of leaves for each subtree. Subtrees with fewer than $20$ leaves are discarded.}
    \label{fig:breaktree}
\end{figure}

    \item  Starting from the fourth iteration, the tolerance values are calculated based on the previously accepted samples. We use the weighted means in the accepted samples from the previous iteration as estimates of the parameters. We then run $10$ simulations with these values, and set the values in the tolerance vector, $\boldsymbol\epsilon_t$, to be $40$ times the standard deviation of the summary statistics from these simulated trees, multiplied by a scaling factor that decreases by a constant multiple if the acceptance rate is above a specified threshold, and is unchanged otherwise; we use $e^{-0.2}$ for the multiplying factor, and 0.03 for the threshold value, as above. If the acceptance rate is above 0.03, we also re-simulate the trees.

\end{itemize}

\paragraph{nLTT statistic}

A single observed tree can give a highly variable nLTT curve. As a result, the nLTT statistic, which is the area between the nLTT curves of the simulated and observed trees, has a high variance, which makes it difficult to predetermine a reasonable tolerance value for the nLTT statistic. Therefore, at the iterations where we consider the nLTT statistic ($t\geq 20$), we first reject or accept $2N$ samples based on the other summary statistics, then select the $N$ samples out of the accepted $2N$ with the smallest nLTT statistic.

\section{Goodness-of-fit analysis for the squamate phylogeny}\label{sec:gof}

\begin{figure}
    \centering
    \includegraphics[width=\linewidth]{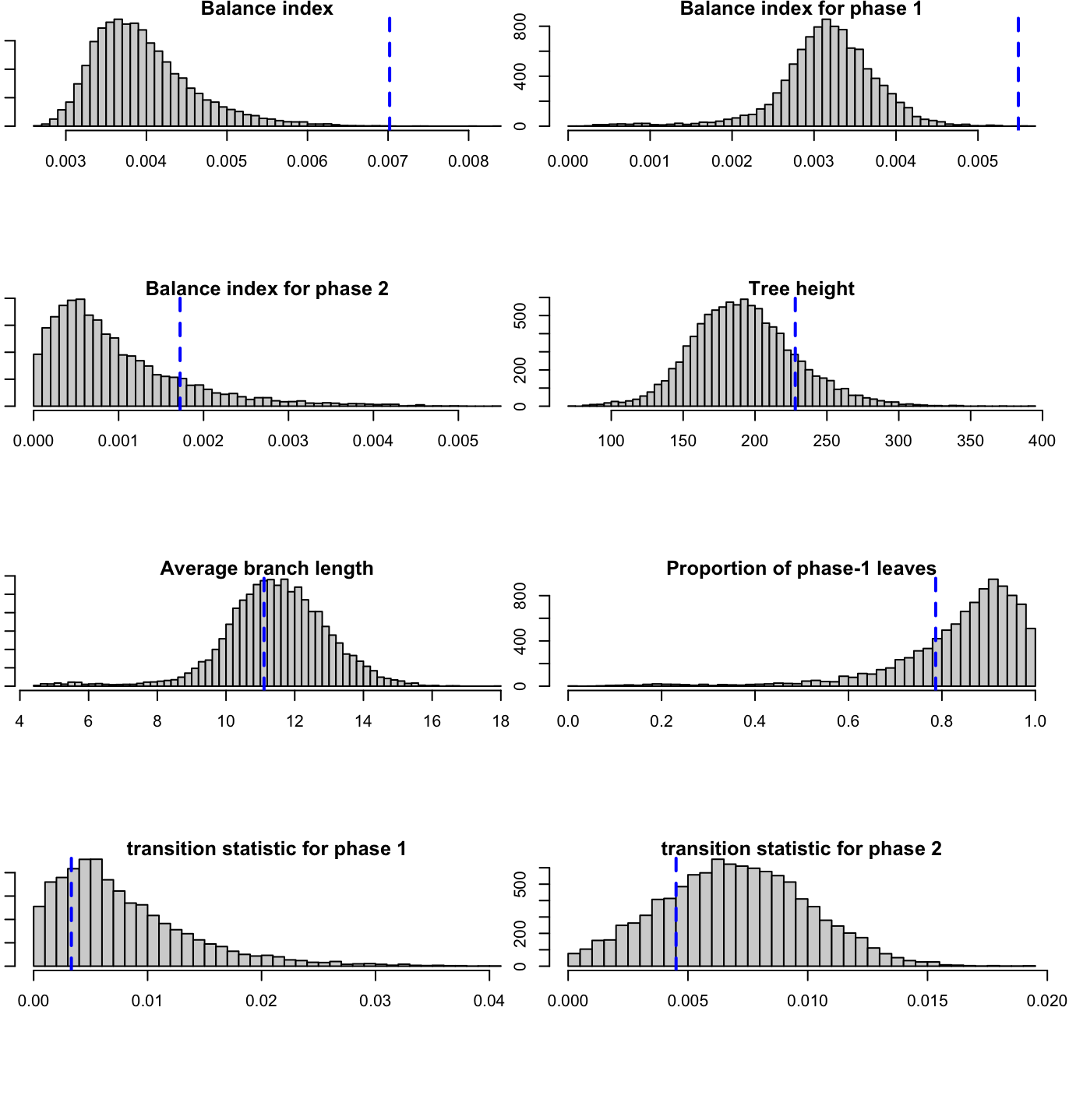}
    \caption{Posterior predictive distributions for each of the summary statistics. The blue dashed line indicates the values of the observed statistics.}
    \label{fig:post_pred_sf2}
\end{figure}

To assess if our inferred posterior parameters fit the observed squamate phylogeny, we sampled from the estimated posterior and constructed posterior predictive distributions of the summary statistics. Since the estimated posterior is not analytically tractable, we generated $100,000$ samples from the kernel density estimate of the final posteriors (after the 30th iteration) using a Markov Chain Monte Carlo run. We discarded the first 1000 samples as burn-in, and further thinned the data by 10-fold. We then plotted the posterior predictive distributions of the summary statistics from our samples. The results are shown in Figure \ref{fig:post_pred_sf2}. We see that for most statistics, the observed values are not extreme for the posterior predictive distribution, indicating a good fit. 
However, for the balance index and balance index in phase 1, the observed values are significantly larger. This suggests that the MBT model does not completely accommodate the imbalance in the observed tree.

\section{ABC SMC model selection}

We perform model selection between a reducible and an irreducible model for the observed squamate phylogeny to investigate the existence of a viviparity-to-oviparity transition (i.e. the transition rate from phase $2$ to phase $1$). We use an ABC sequential Monte Carlo (SMC) model selection method \citep{toni2009approximate}. For comparison purposes, we assume both models start in phase $1$. 

For inference with a single observed tree, the tolerance values cannot be computed directly from the observed dataset. Instead, the tolerance values for each summary statistic (from the fourth iteration) are computed from simulations. Therefore, we use a slightly modified version of the ABC SMC model. To determine the tolerance vector, we compute the tolerance vector under each model, $\boldsymbol{\epsilon}^{M_1}$, $\boldsymbol{\epsilon}^{M_2}$. At iteration $t$, we set the tolerance vector to be $\boldsymbol{\epsilon}_t=\alpha_t\text{max}(\boldsymbol{\epsilon}_t^{M_1},\boldsymbol{\epsilon}_t^{M_2})$, where $\alpha_t$ is the scaling factor calculated based on the acceptance rate. The rest of the algorithm follows the ABC SMC algorithm. In the algorithm, we propose parameters from two models (here, the reducible and irreducible models), weighted by a prior distribution that we set to be $\frac12$ for each model. For iterations beyond the first, the prior distribution for each model is constructed as before, but only from the accepted samples from that model from the previous iteration. The proportions of accepted samples that belong to either model then give us posterior probabilities for the models.

\newpage

\section{Supplementary Figures and Tables}

\noindent\textbf{\hypertarget{text:4ptrace}{Supplementary Figure S11}. Trace plots of posterior means and credible intervals at each iteration for the reducible case with equal death rates.} The approximate posterior means for the $\mathbf{(a)}$ phase-1 birth rate $\lambda_1$ (blue), phase-2 birth rate $\lambda_2$ (orange), death rate $\mu$ (green), and transition rate from phase $1$ to phase $2$, $q_{12}$ (red); $\mathbf{(b)}$ growth rate $\omega$. Error bars represent the 50\% credible intervals of the approximate posterior distribution. Horizontal dashed lines indicate the true parameter values. Vertical dotted lines indicate the iterations where the tolerance values were decreased.

\begin{figure}[htb]\centering
\sidesubfloat[]{\includegraphics[width=0.9\textwidth]{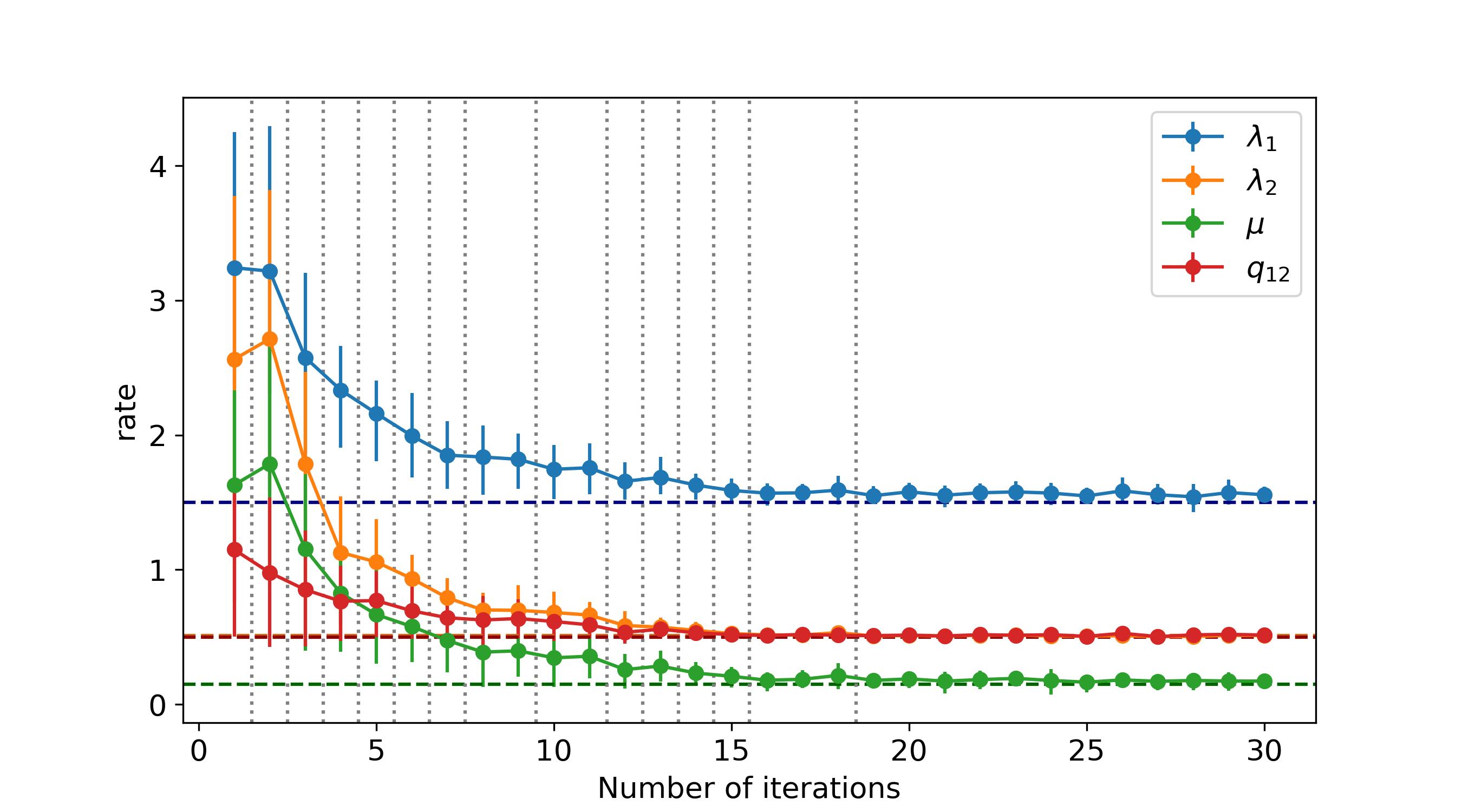}\label{fig:4pmodmean}}
\hfill
\sidesubfloat[]{\includegraphics[width=0.9\textwidth]{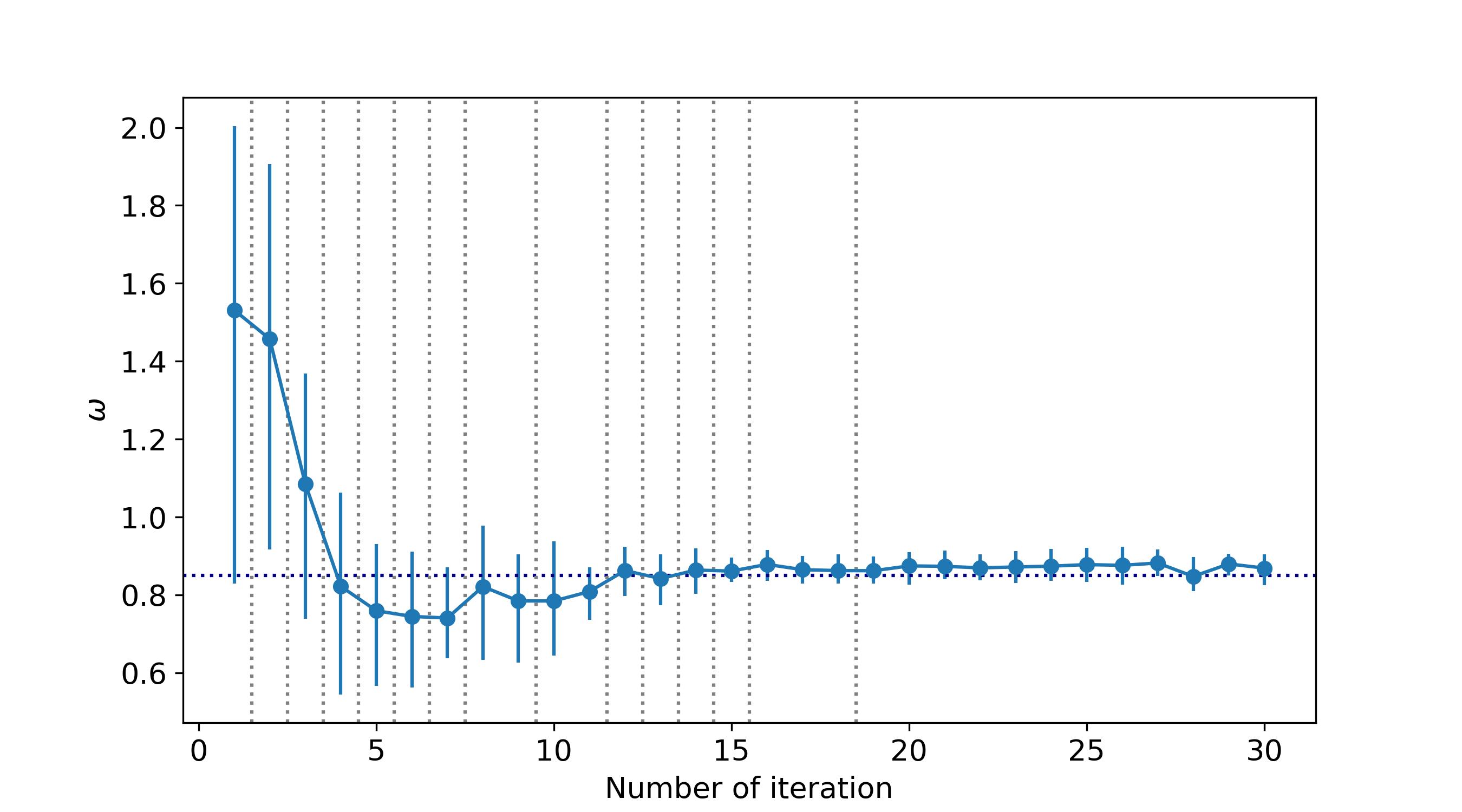}\label{fig:4pmoddom}}
\end{figure}

\newpage

\noindent\textbf{\hypertarget{text:4ppostmodall1}{Supplementary Figure S12}. The final posterior distributions for the reducible case with equal death rates for one run of the ABC-PMC algorithm.}
\begin{figure}[htp]
 \centering
		\includegraphics[width=\textwidth]{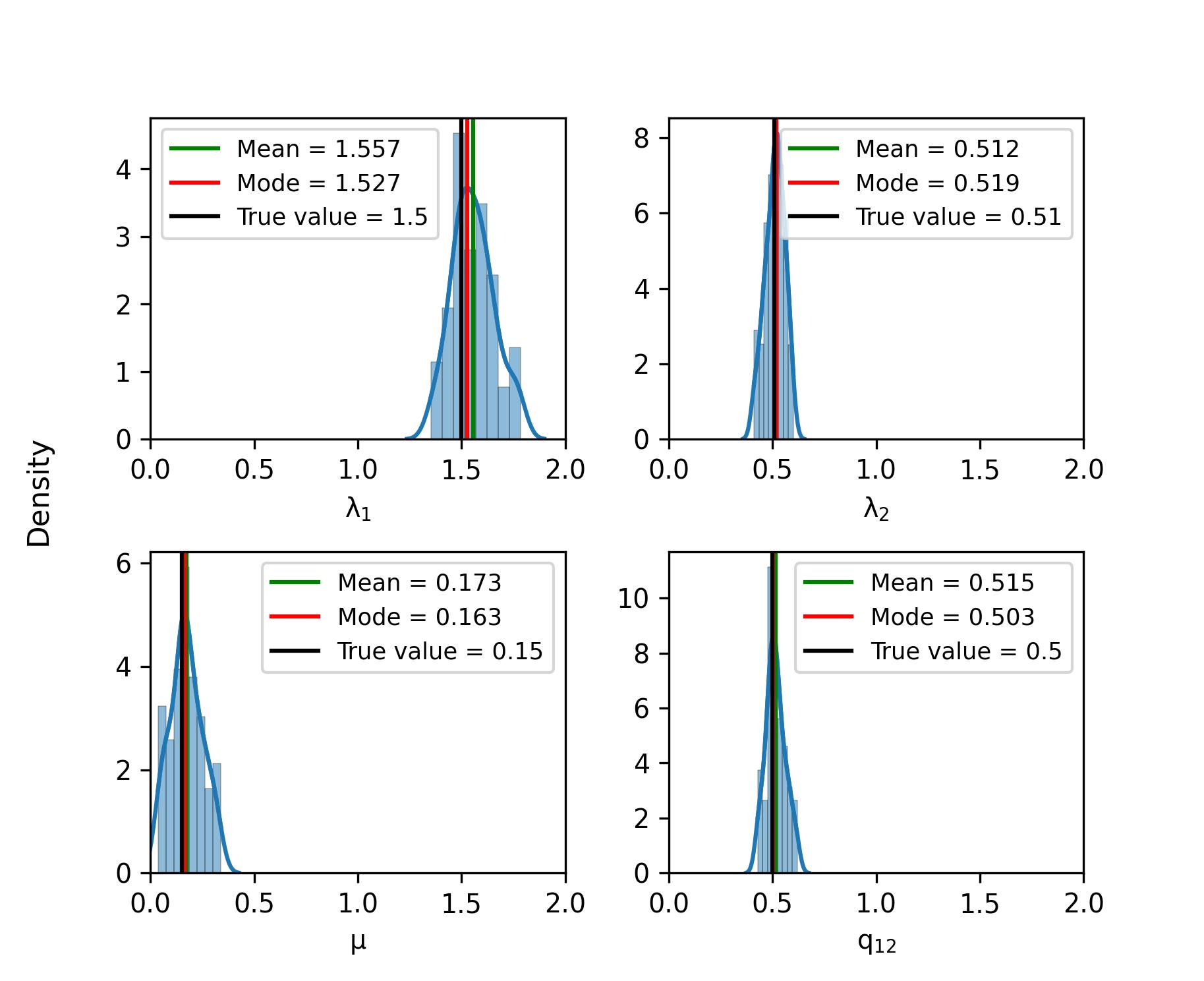}
\end{figure}

\newpage
    
\noindent\textbf{\hypertarget{text:5ptrace}{Supplementary Figure S13}. Trace plots of posterior means and credible intervals at each iteration for the reducible case with arbitrary death rates.} The approximate posterior means for the $\mathbf{(a)}$ phase-1 birth rate $\lambda_1$ (blue), phase-1 death rate $\mu_1$ (orange); $\mathbf{(b)}$ phase-2 birth rate $\lambda_2$ (blue), phase-2 death rate $\mu_2$ (orange); $\mathbf{(c)}$ transition rate from phase $1$ to phase $2$, $q_{12}$; $\mathbf{(d)}$ growth rate $\omega$. Error bars represent the 50\% credible intervals of the approximate posterior distribution. Horizontal dashed lines indicate the true parameter values. Vertical dotted lines indicate the iterations where the tolerance values were decreased.

\begin{figure}[htb]\centering
\sidesubfloat[]{\includegraphics[width=0.5\textwidth]{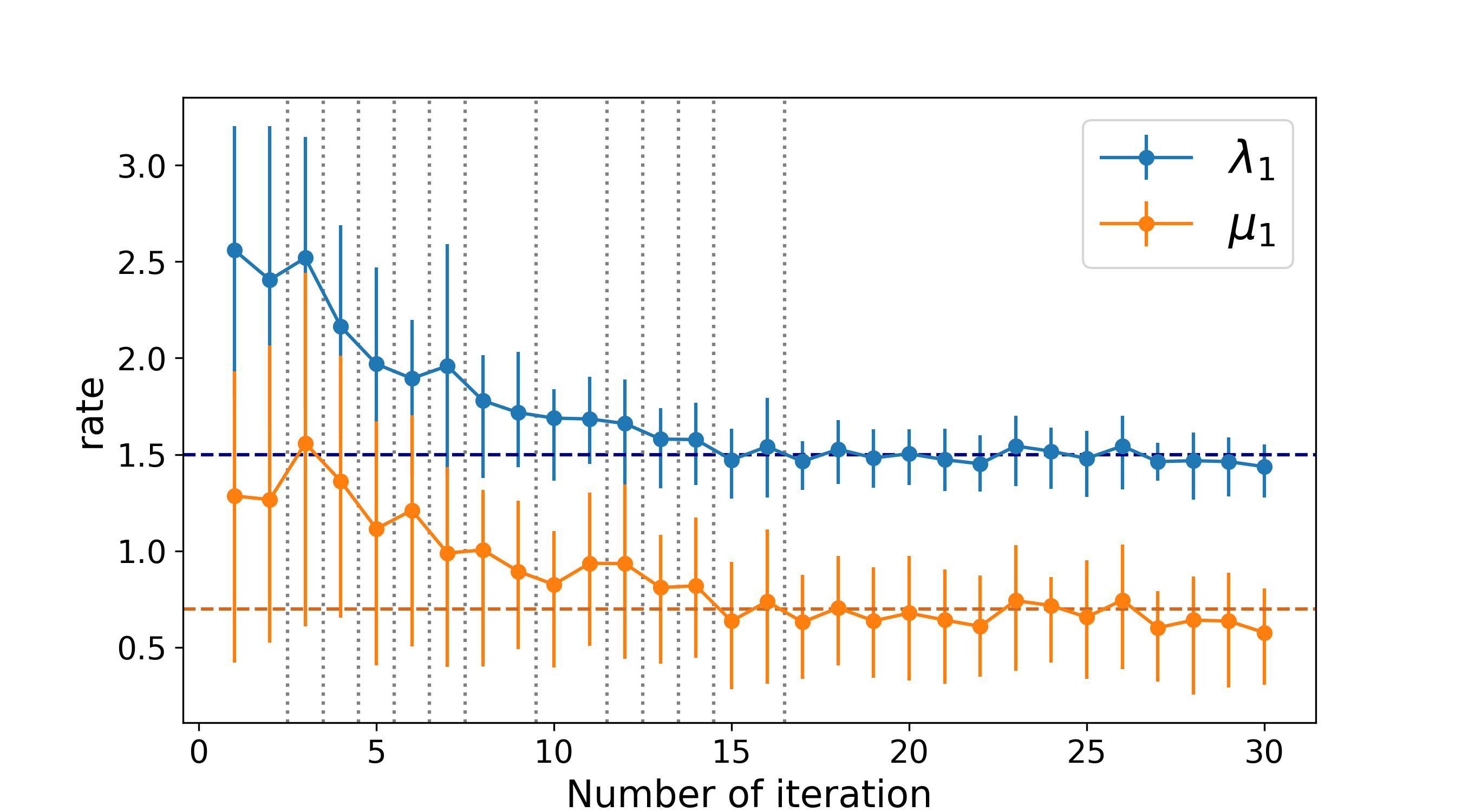}}
\sidesubfloat[]{\includegraphics[width=0.5\textwidth]{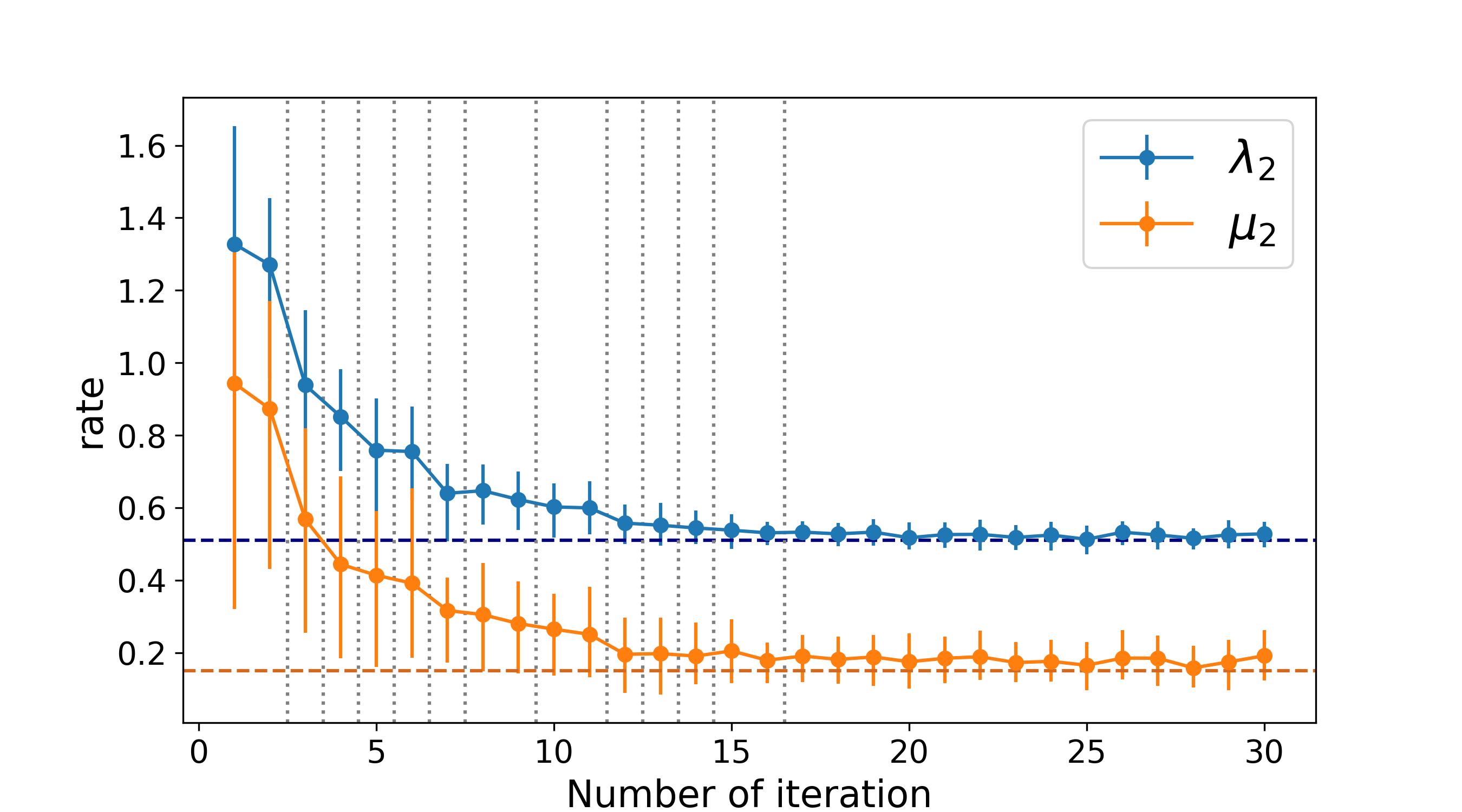}}
\hfil
\sidesubfloat[]{\includegraphics[width=0.5\textwidth]{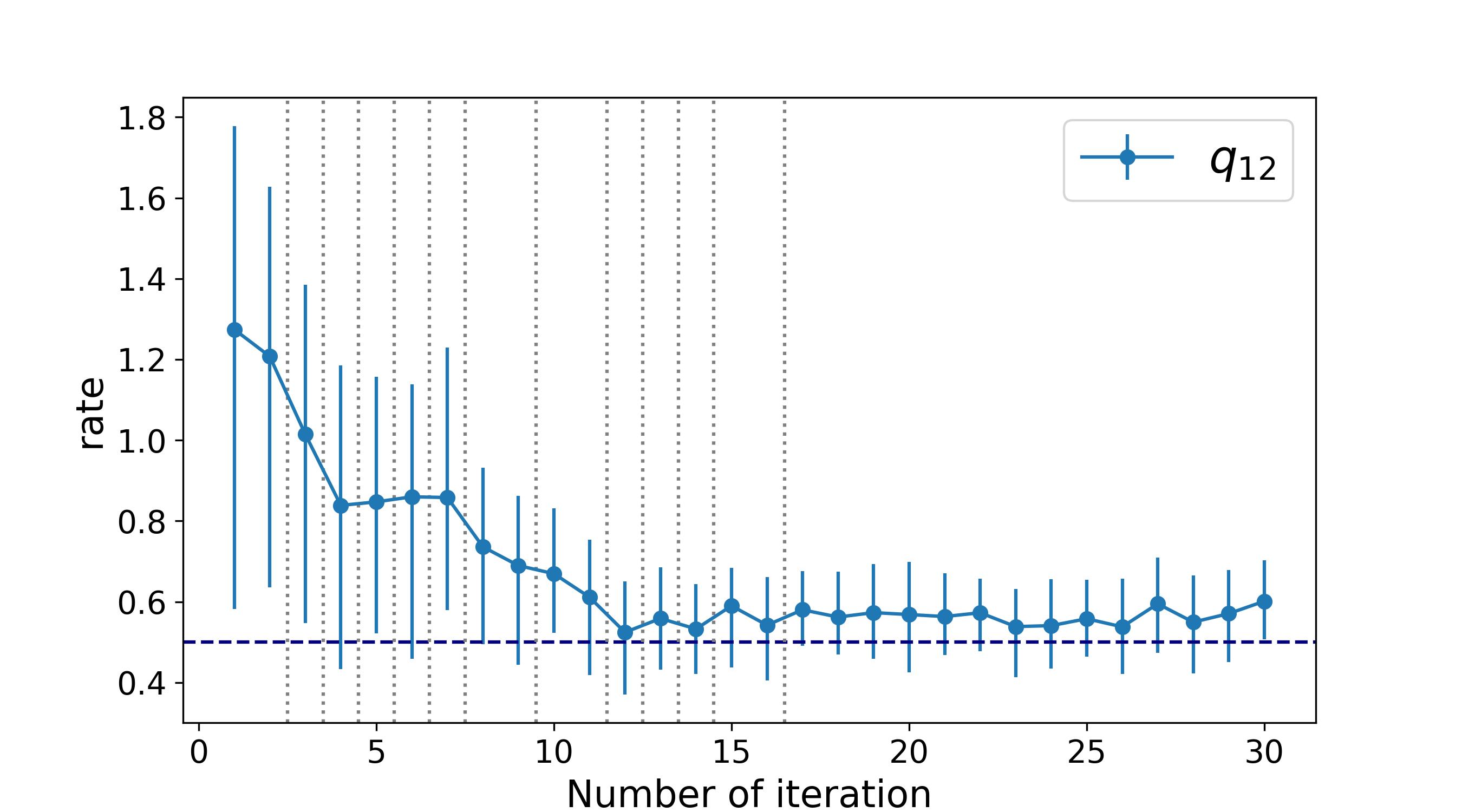}}
\sidesubfloat[]{\includegraphics[width=0.5\textwidth]{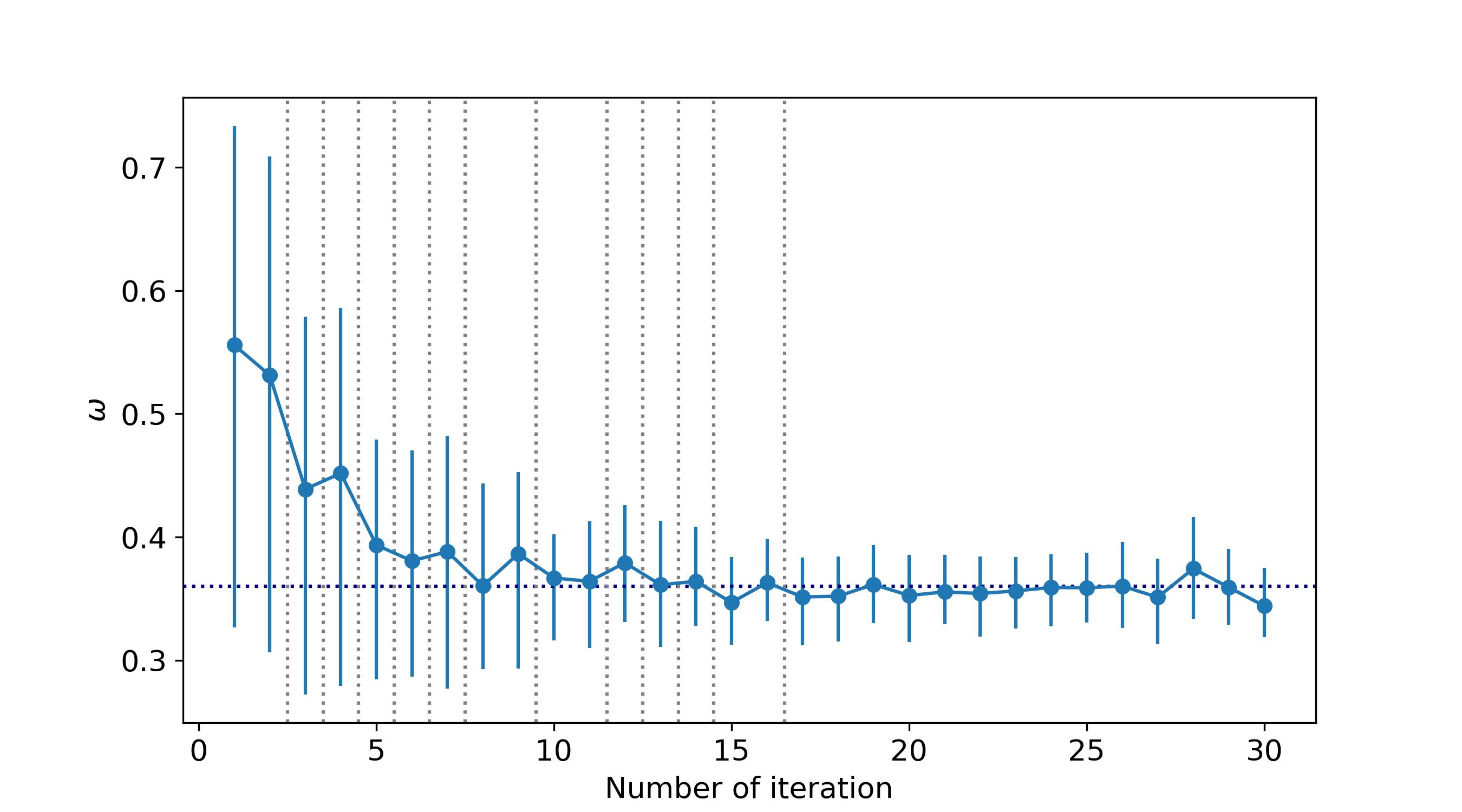}\label{fig:5pgrowth}}

\label{fig:5pmean}
\end{figure}

\newpage

\noindent\textbf{Supplementary Figure S14.\label{text:posterior5p} The final posterior distributions for the reducible case with arbitrary death rates for one run of the ABC-PMC algorithm.}

\begin{figure}[htb]\centering
	\includegraphics[width=0.8\textwidth]{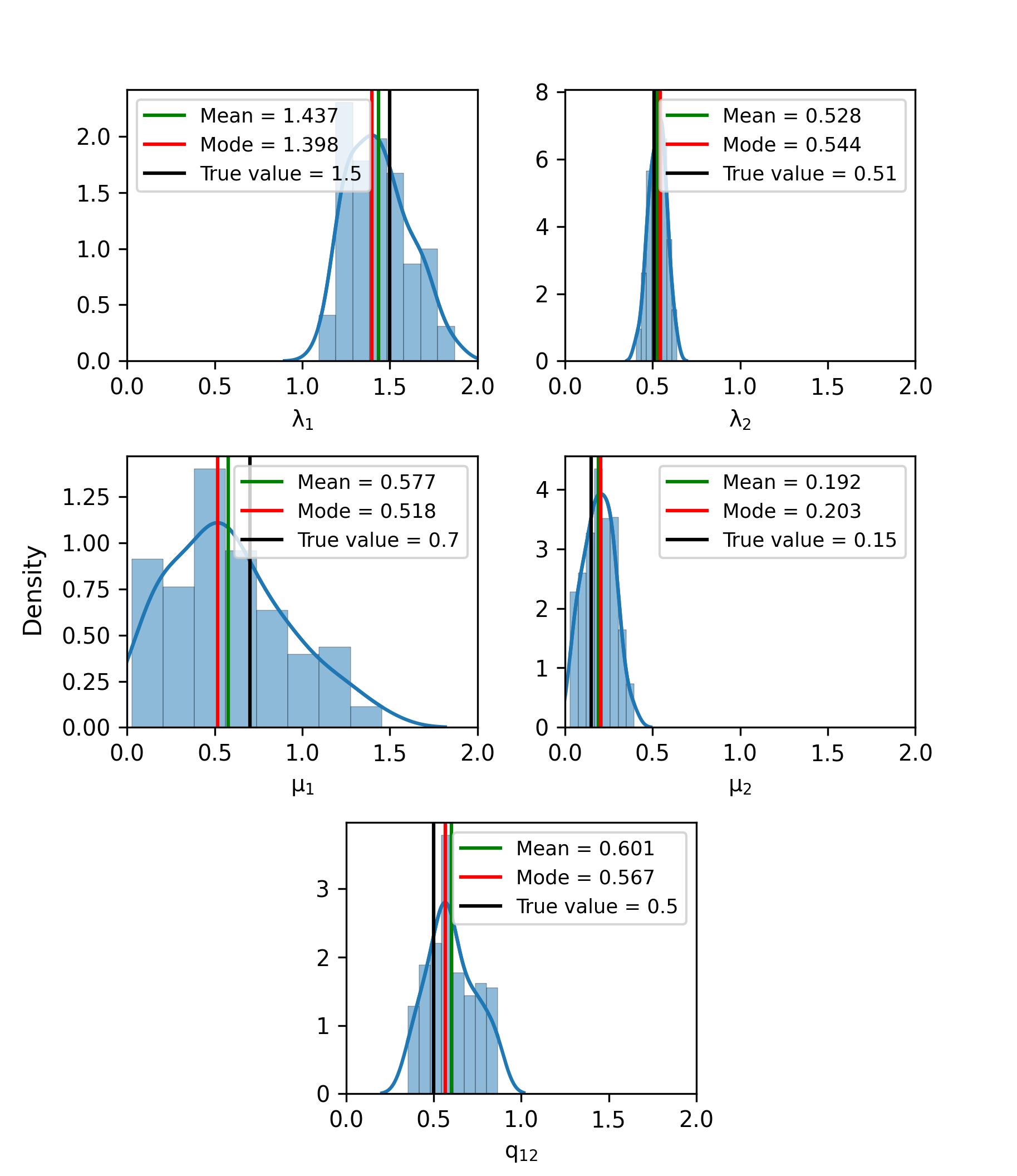}
\end{figure}

\newpage

\noindent\textbf{\hypertarget{text:6ptrace}{Supplementary Figure S15}. Trace plots of posterior means and credible intervals at each iteration for the irreducible case.} The approximate posterior means for the $\mathbf{(a)}$ phase-1 birth rate $\lambda_1$ (blue), phase-1 death rate $\mu_1$ (orange); $\mathbf{(b)}$ phase-2 birth rate $\lambda_2$ (blue), phase-2 death rate $\mu_2$ (orange); $\mathbf{(c)}$ transition rate from phase $1$ to phase $2$, $q_{12}$ (blue), transition rate from phase $2$ to phase $1$, $q_{21}$ (orange); $\mathbf{(d)}$ growth rate $\omega$. Error bars represent the 50\% credible intervals of the approximate posterior distribution. Horizontal dashed lines indicate the true parameter values. Vertical dotted lines indicate the iterations where the tolerance values were decreased.

\begin{figure}[htb]\centering
\sidesubfloat[]{\includegraphics[width=0.5\textwidth]{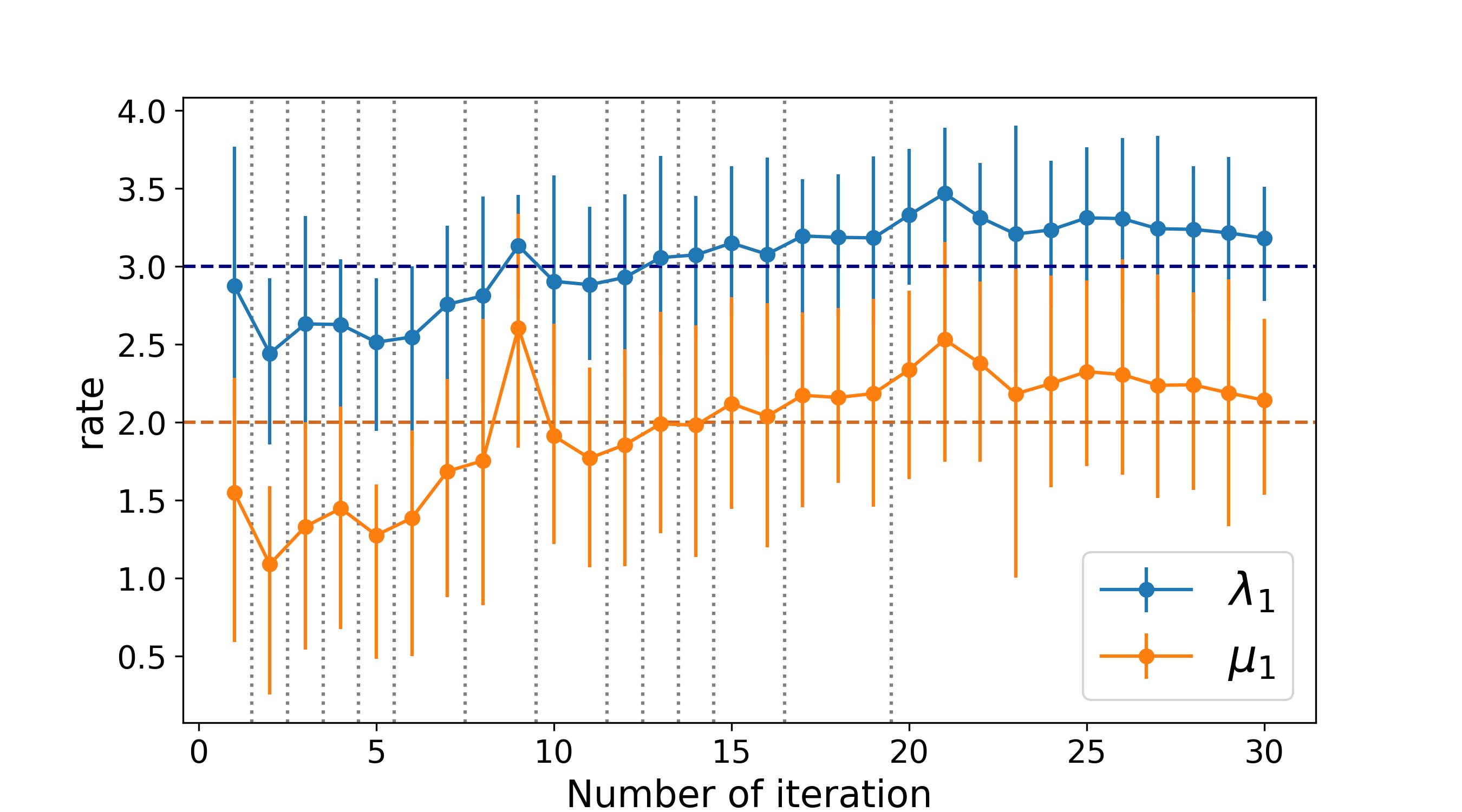}}
\sidesubfloat[]{\includegraphics[width=0.5\textwidth]{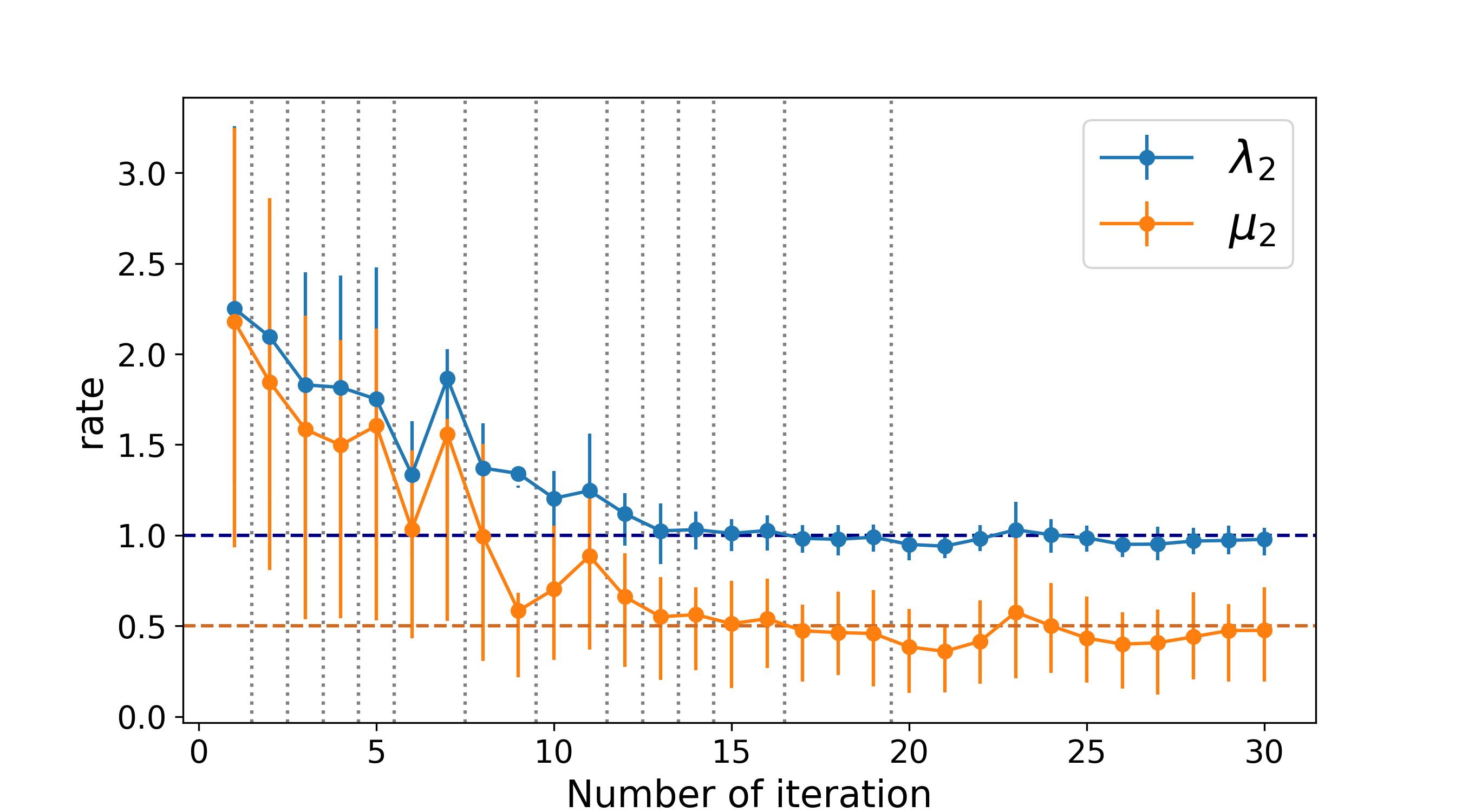}}
\hfil
\sidesubfloat[]{\includegraphics[width=0.5\textwidth]{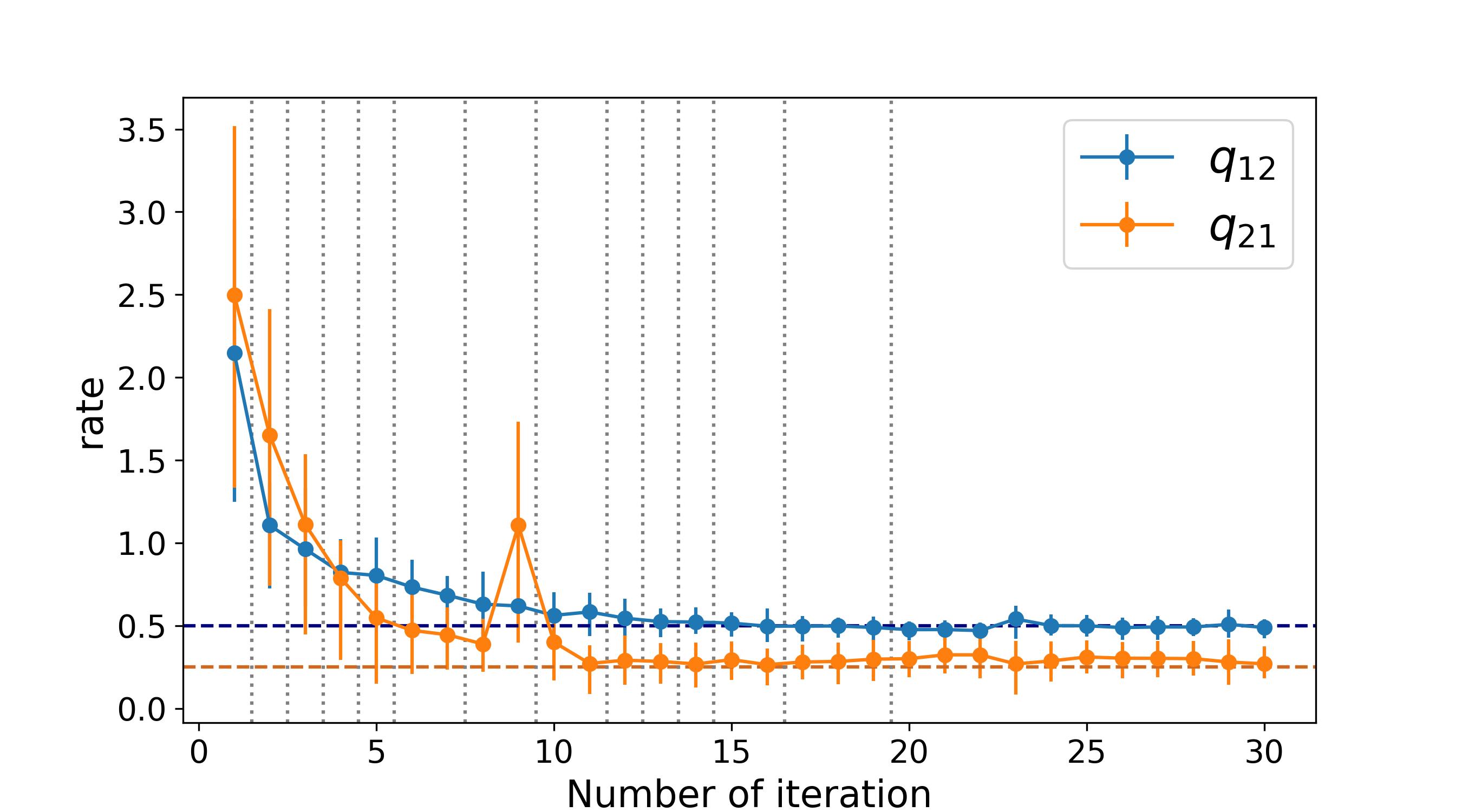}}
\sidesubfloat[]{\includegraphics[width=0.5\textwidth]{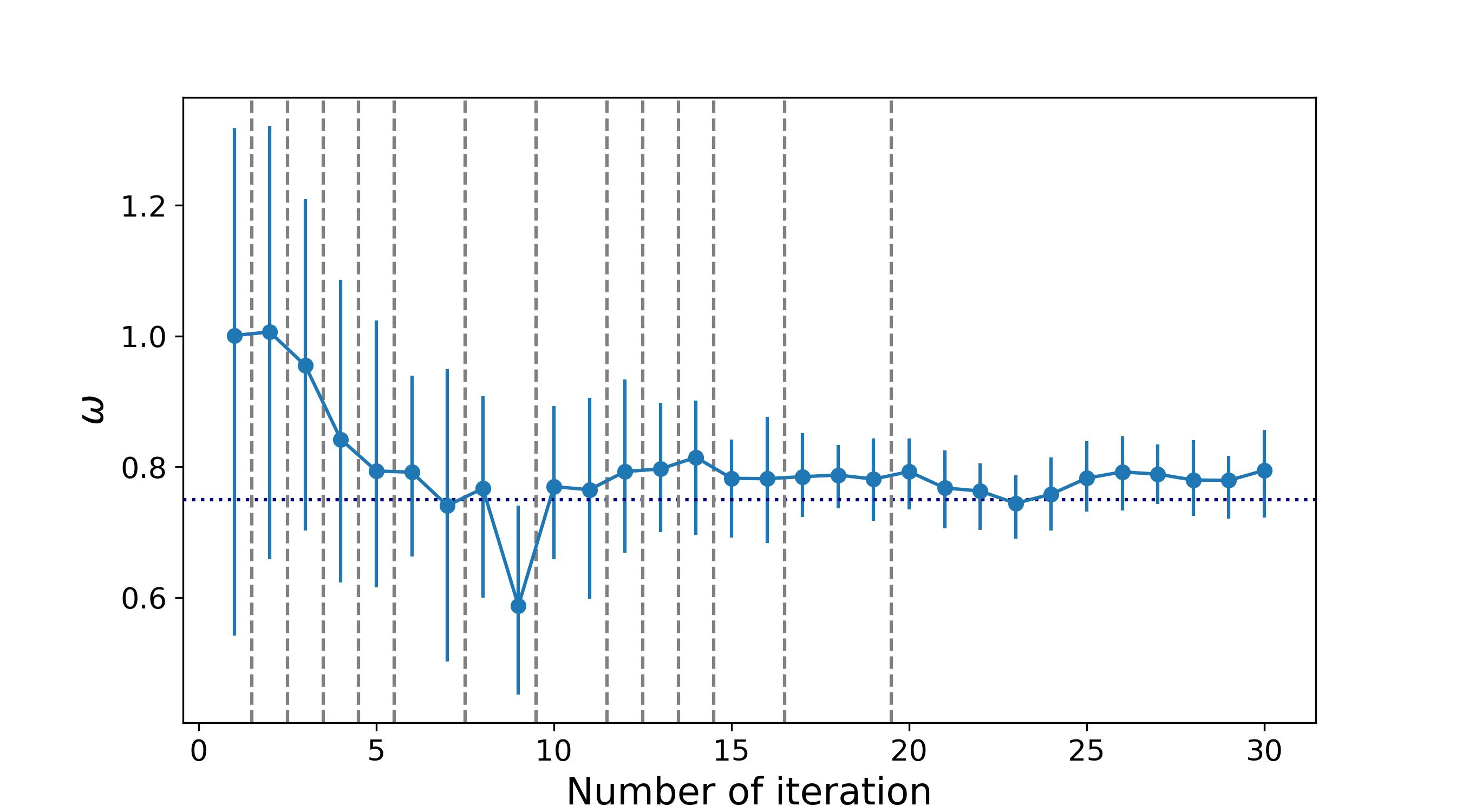}}
\end{figure}

\newpage

\noindent\textbf{\hypertarget{text:posterior6p}{Supplementary Figure S16}. The final posterior distributions for the irreducible case for one run of the ABC-PMC algorithm.}
\begin{figure}[htb]\centering
    \includegraphics[width=0.8\textwidth]{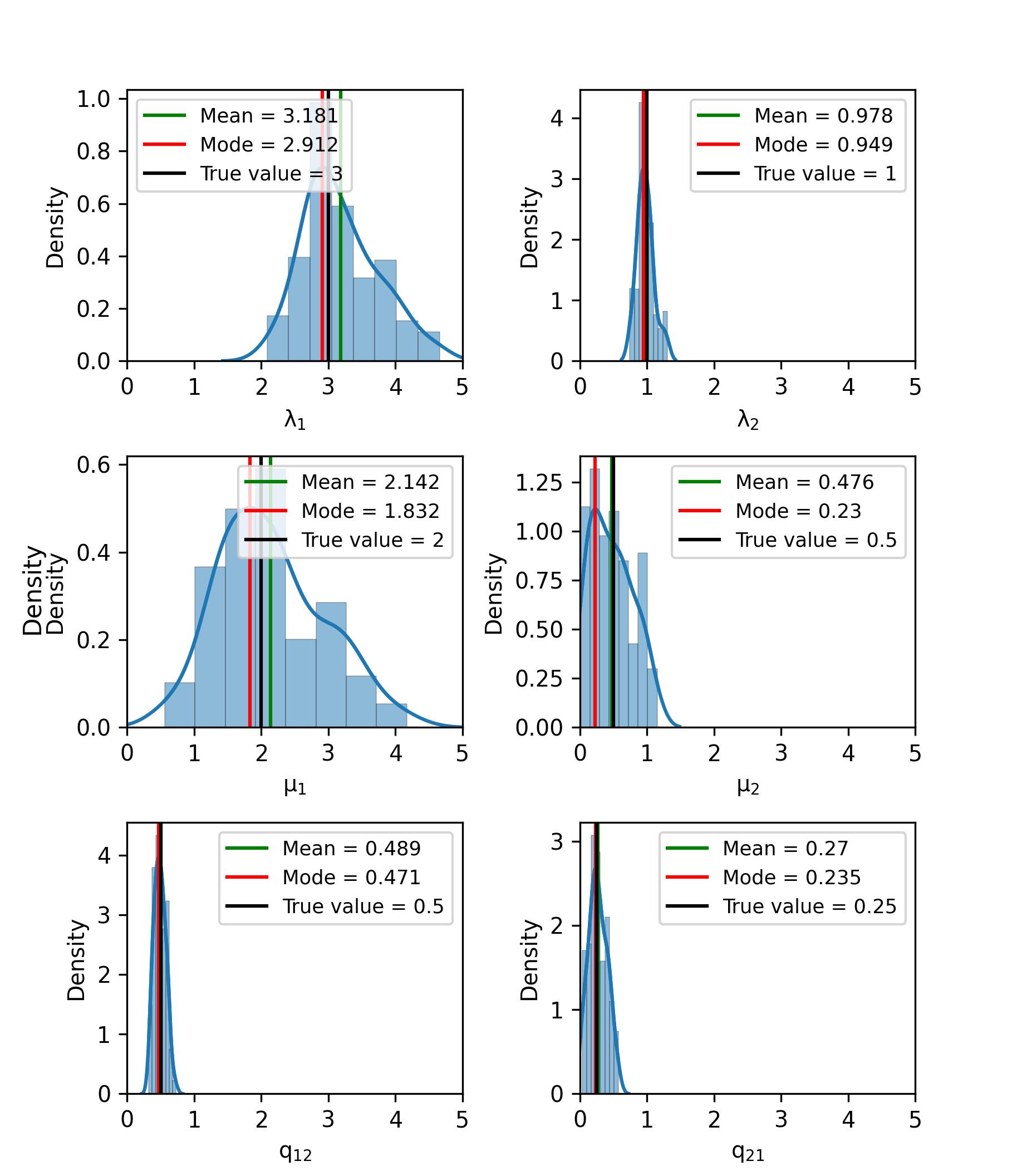}
\end{figure}

\newpage

\noindent\textbf{\hypertarget{text:size}{Supplementary Figure S17}. Parameter estimation for different tree sizes with 25 observed trees.}
\begin{figure}[htb]\centering
    \includegraphics[width=0.85\textwidth]{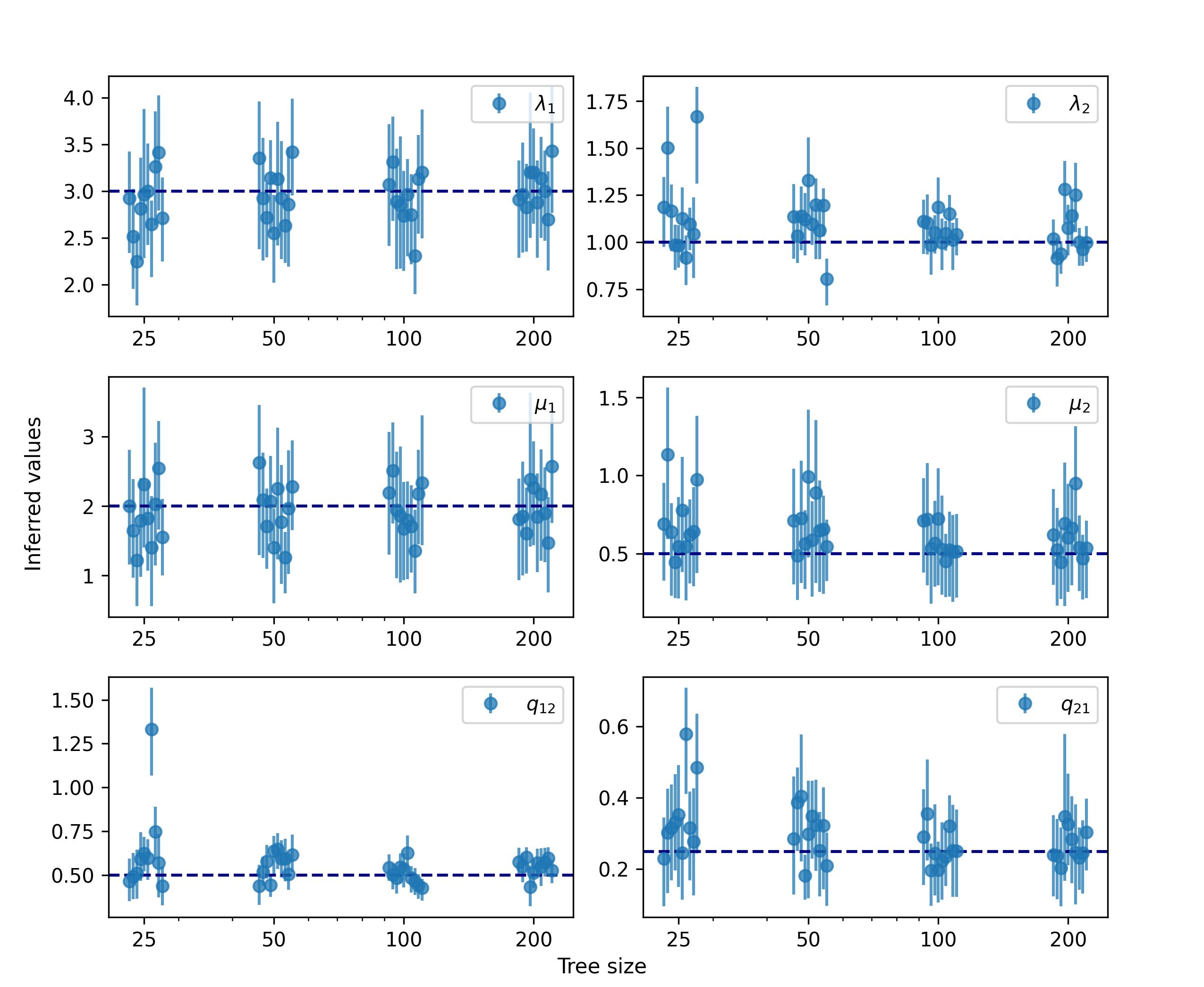}
\end{figure}

\newpage

\noindent\textbf{\hypertarget{text:size2}{Supplementary Figure S18}. Parameter estimation for different tree sizes with the total number of leaves fixed at 5000.}
\begin{figure}[htb]\centering
    \includegraphics[width=0.85\textwidth]{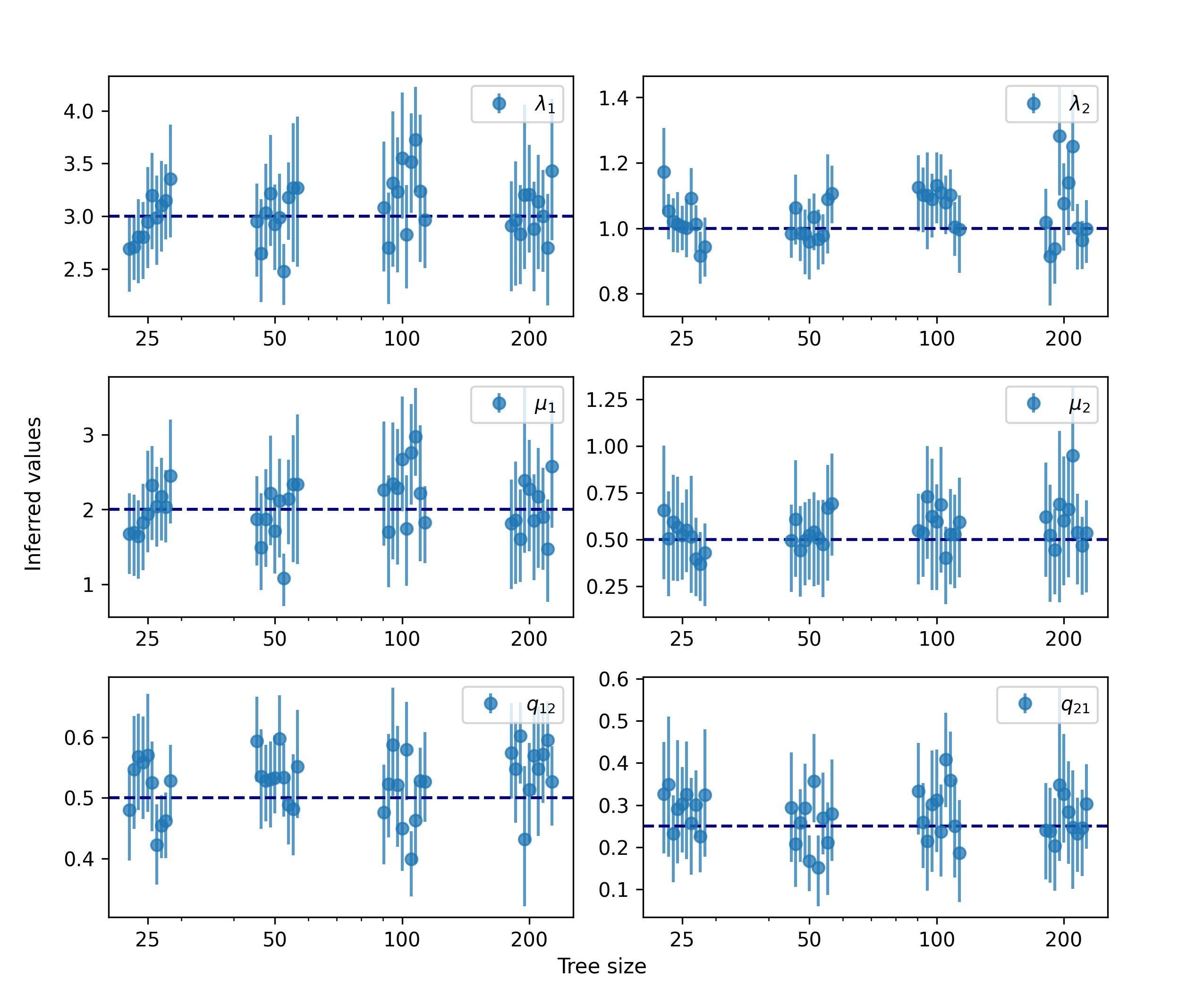}
\end{figure}

\newpage

\noindent\textbf{\hypertarget{text:e_large_d_comp}{Supplementary Figure S19}. Parameter inference error for ABC and ML methods for the irreducible case, for the default parameters for 50 replicates. Each dataset has one tree with 5000 leaves.}
\begin{figure}[htb]\centering
    \includegraphics[height=10cm]{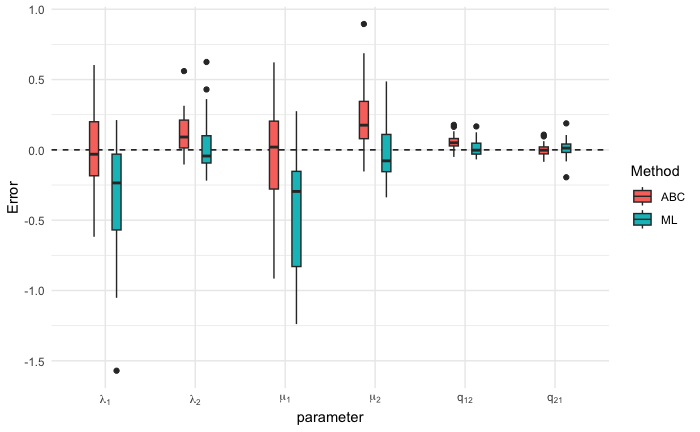}
\end{figure}

\newpage

\noindent\textbf{\hypertarget{text:large_diff_compare}{Supplementary Figure S20}.Inferred vs true parameter values for ABC and ML methods for the irreducible case, with varying parameters for 100 replicates. Each dataset has one tree with 5000 leaves.}
\begin{figure}[htb]\centering
\hspace{-2cm}
\begin{tabular}{cc}
  \includegraphics[width=75mm]{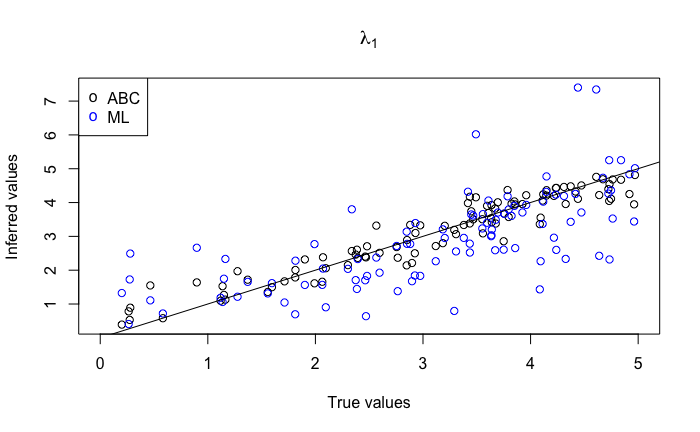} &   \includegraphics[width=75mm]{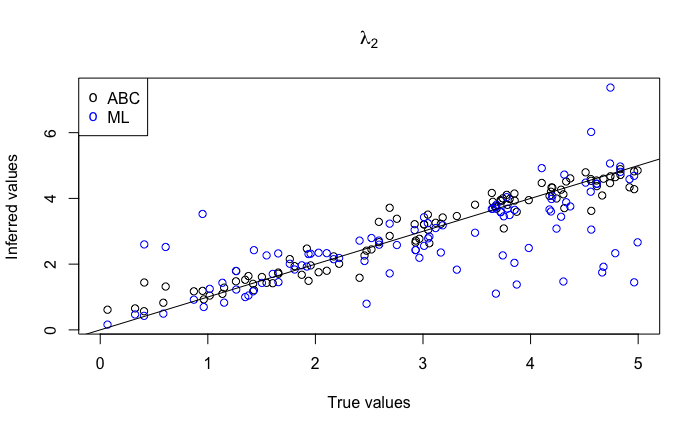}\\
 \includegraphics[width=75mm]{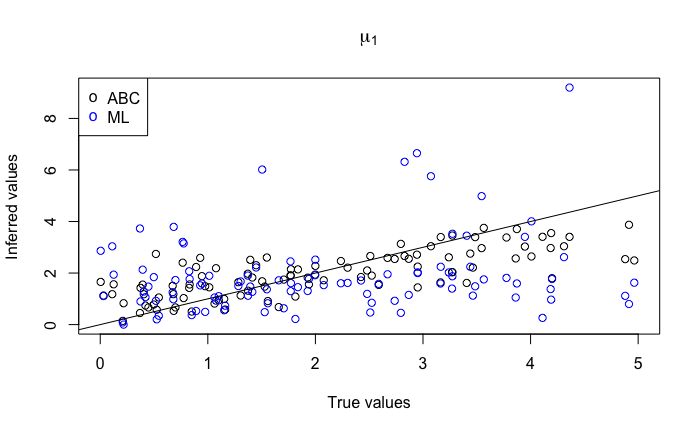} &   \includegraphics[width=75mm]{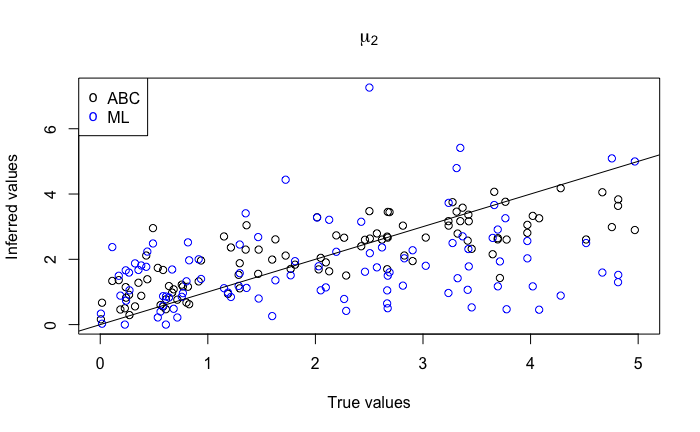} \\
 \includegraphics[width=75mm]{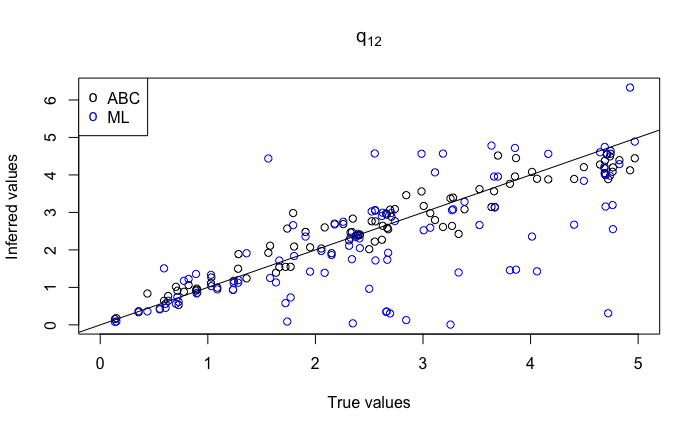} &   \includegraphics[width=75mm]{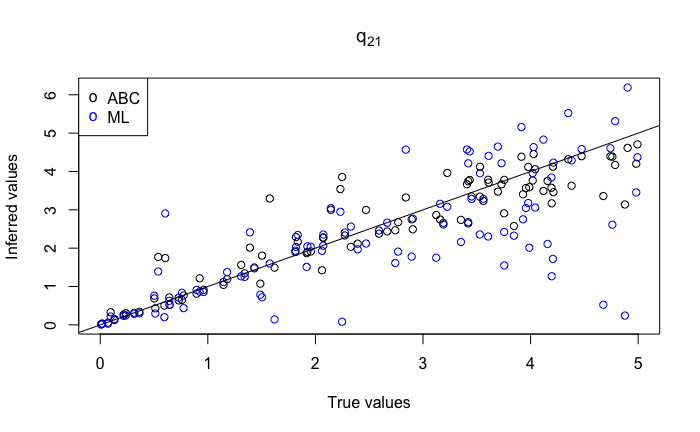} 
\end{tabular}
\end{figure}

\newpage

\noindent\textbf{\hypertarget{text:paritytrace}{Supplementary Figure S21}. Trace plots of posterior means and credible intervals at each iteration for the squamata data under the irreducible model where processes start in phase $2$} The approximate posterior means for the $\mathbf{(a)}$ phase-1 birth rate $\lambda_1$ (blue), phase-1 death rate $\mu_1$ (orange); $\mathbf{(b)}$ phase-2 birth rate $\lambda_2$ (blue), phase-2 death rate $\mu_2$ (orange); $\mathbf{(c)}$ transition rate from phase $1$ to phase $2$, $q_{12}$ (blue), transition rate from phase $2$ to phase $1$, $q_{21}$ (orange); $\mathbf{(d)}$ growth rate $\omega$. 

\begin{figure}[htb]\centering
\sidesubfloat[]{\includegraphics[width=0.5\textwidth]{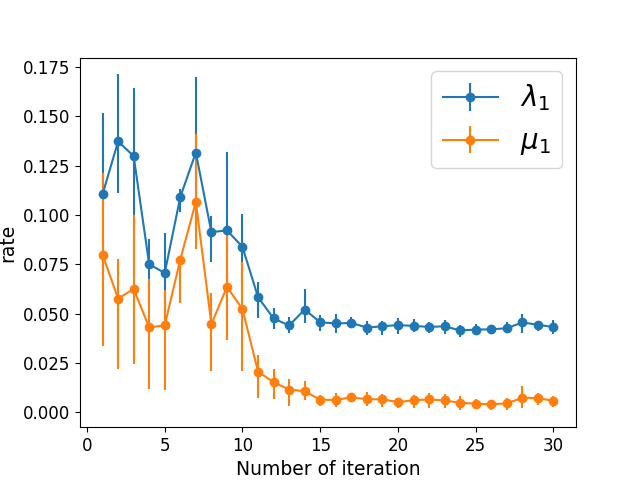}}
\sidesubfloat[]{\includegraphics[width=0.5\textwidth]{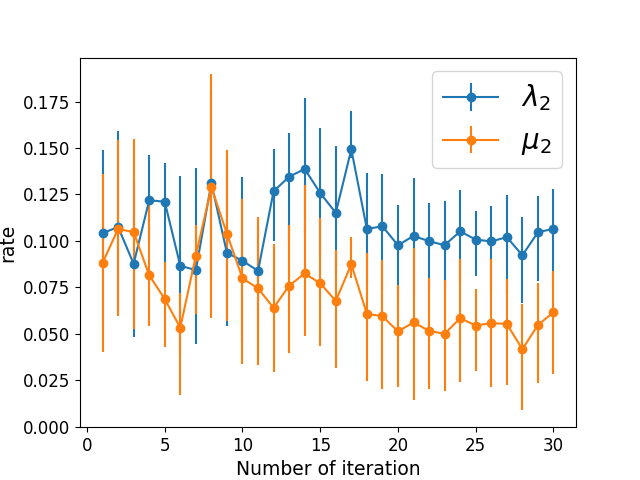}}
\hfil
\sidesubfloat[]{\includegraphics[width=0.5\textwidth]{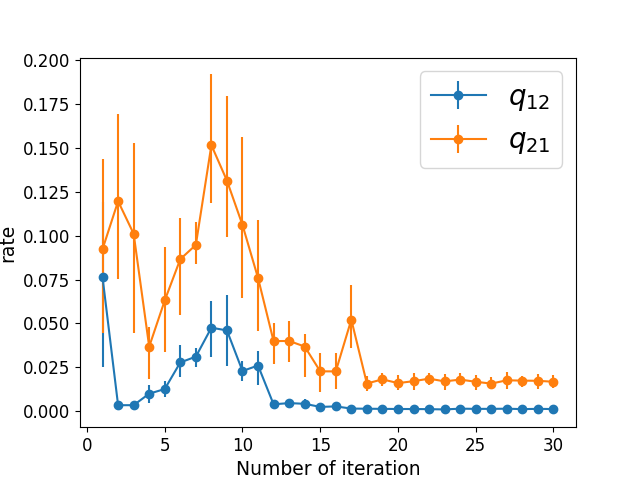}}
\sidesubfloat[]{\includegraphics[width=0.5\textwidth]{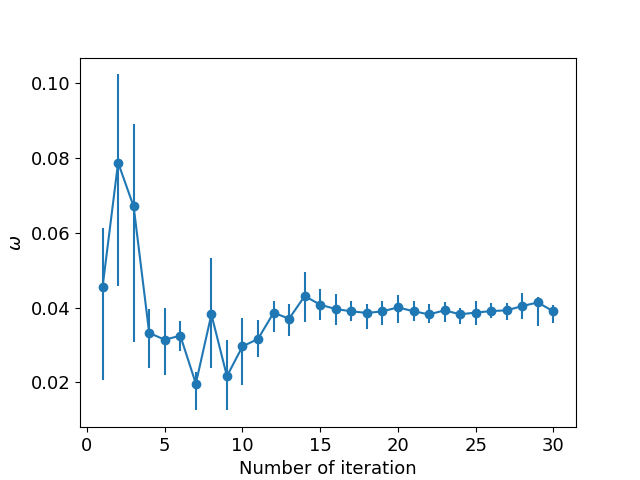}}
\label{fig:parity_trace}
\end{figure}

\newpage

\noindent\textbf{\hypertarget{table:rrmse_default_s5000}{Supplementary Table S1}. RRMSE for ABC and ML methods for the irreducible case, for the default parameters for $50$ replicates. Each dataset consists a single tree with $5000$ leaves. } 

\begin{table}[htb]
\centering
 \begin{tabular}{|l|c|c|c|c|c|c|c|} 
 \hline
 RRMSE & $\lambda_1$ & $\lambda_2$ & $\mu_1$ & $\mu_2$ & $q_{12}$ & $q_{21}$\\ [0.5ex] 
 \hline\hline
ABC & 0.08 & 0.17 & 0.16 & 0.59 & 0.15 & 0.16\\
ML & 0.17& 0.17& 0.30& 0.40 & 0.12 & 0.23\\ [1ex] 
 \hline
 \end{tabular}
 
\end{table}

\noindent\textbf{\hypertarget{table:rrmse_diff_s5000}{Supplementary Table S2}. RRMSE for ABC and ML methods for the irreducible case, with varying parameters for $100$ replicates. Each dataset consists a single tree with $5000$ leaves. }

\begin{table}[htb]
\centering
 \begin{tabular}{|l|c|c|c|c|c|c|c|} 
 \hline
 RRMSE & $\lambda_1$ & $\lambda_2$ & $\mu_1$ & $\mu_2$ & $q_{12}$ & $q_{21}$\\ [0.5ex] 
 \hline\hline
    ABC & 0.12 & 0.11 & 0.44 & 0.41 & 0.15 & 0.21\\
    ML & 0.33 & 0.36 & 0.86 & 0.75 & 0.46 & 0.43\\ [1ex] 
 \hline
 \end{tabular}
\end{table}


\end{document}